\documentclass[12pt,preprint]{elsarticle}
\usepackage[latin9]{inputenc}
\usepackage{units}
\usepackage{mathrsfs}
\usepackage{amsmath}
\usepackage{amssymb}
\usepackage{graphicx}
\usepackage{esint}

\makeatletter

\providecommand{\tabularnewline}{\\}

\RequirePackage{lineno}

\journal{Journal of Computational Physics}

\makeatother

\begin{document}

\title{Numerical stability of relativistic beam multidimensional PIC simulations
employing the Esirkepov algorithm}

\author{Brendan B. Godfrey}

\address{University of Maryland, College Park, Maryland 20742, USA}

\author{Jean-Luc Vay}

\address{Lawrence Berkeley National Laboratory, Berkeley, California 94720,
USA}
\begin{abstract}
Rapidly growing numerical instabilities routinely occur in multidimensional
particle-in-cell computer simulations of plasma-based particle accelerators,
astrophysical phenomena, and relativistic charged particle beams.
Reducing instability growth to acceptable levels has necessitated
higher resolution grids, high-order field solvers, current filtering,
\emph{etc.} except for certain ratios of the time step to the axial
cell size, for which numerical growth rates and saturation levels
are reduced substantially. This paper derives and solves the cold
beam dispersion relation for numerical instabilities in multidimensional,
relativistic, electromagnetic particle--in-cell programs employing
either the standard or the Cole-Karkkainnen finite difference field
solver on a staggered mesh and the common Esirkepov current-gathering
algorithm. Good overall agreement is achieved with previously reported
results of the WARP code. In particular, the existence of select time
steps for which instabilities are minimized is explained. Additionally,
an alternative field interpolation algorithm is proposed for which
instabilities are almost completely eliminated for a particular time
step in ultra-relativistic simulations.\end{abstract}
\maketitle

\section{Introduction}

In a laser plasma accelerator (LPA), a laser pulse is propagated through
a plasma, creating a wake of very strong electric fields of alternating
polarity \citep{TajimaPRL79}. An electron beam injected with the
appropriate phase can thus be accelerated to high energy in a distance
much shorter than those for conventional acceleration techniques \citep{LeemansNature06}.
Simulation of a LPA stage from first principles using the Particle-In-Cell
(PIC) technique in the laboratory frame is very demanding computationally,
as the evolution of micron-scale laser oscillations needs to be followed
over millions of time steps as the laser pulse propagates through
a meter-long plasma for a 10 GeV stage \citep{BruhwilerAAC08}. 

A method recently was demonstrated to speed up full PIC simulations
of a certain class of relativistic interactions by performing the
calculation in a Lorentz boosted frame \citep{VayPRL07}, taking advantage
of the properties of space-time contraction and dilation in special
relativity to render space and time scales (which are separated by
orders of magnitude in the laboratory frame) comparable in a Lorentz
boosted frame, resulting in far fewer computer operations. In the
laboratory frame the laser pulse is much shorter than the wake, whose
wavelength is also much shorter than the acceleration distance ($\lambda_{laser}\ll\lambda_{wake}\ll\lambda_{acceleration}$).
In a Lorentz boosted frame co-propagating with the laser at a speed
near the speed of light, the laser is Lorentz expanded (by a factor
$\left(1+v_{f}\right)\gamma_{f}$, where $\gamma_{f}=\left(1-v_{f}^{2}\right)^{-1/2}$,
$v_{f}$ is the velocity of the frame, normalized to the speed of
light). The plasma (now moving opposite to the incoming laser at velocity
$-v_{f}$) is Lorentz contracted by a factor $\gamma_{f}$. In a boosted
frame moving with the wake (\emph{i.e.}, $\gamma_{f}\approx\gamma_{wake})$,
the laser wavelength, the wake, and the acceleration length are comparable
($\lambda_{laser}<\lambda_{wake}\approx\lambda_{acceleration}$),
leading to far fewer time steps by a factor $\left(1+v_{f}\right)^{2}\gamma_{f}^{2}$,
hence far fewer computer operations \citep{VayPRL07,VayPoP2011}.

A violent numerical instability, associated with the plasma back-streaming
at relativistic velocity $-v_{f}$ in the computational frame, limited
early attempts to apply this method to speedups ranging between two
and three orders of magnitude \citep{BruhwilerAAC08,VayPAC09,MartinsNaturePhysics10}.
Control of the instability was obtained via the combination of: (i)
the use of a tunable electromagnetic solver and an efficient wide-band
digital filtering method \citep{VayJCP2011}, (ii) observation of
the benefits of hyperbolic rotation of space-time on the laser spectrum
in boosted frame simulations \citep{VayPoPL2011}, and (iii) identification
of a special time step at which the growth rate of the instability
is greatly reduced \citep{VayJCP2011}. The combination of these methods
enabled the demonstration of speedups of over a million times \citep{VayPoPL2011}.
The instability is described in some detail in \citep{VayJCP2011}. 

In this paper, the analysis first reported in \citep{godfrey1974numerical},
which introduced the concept of numerical Cherenkov instabilities,
is generalized and extended to two dimensions. (Extension to three
dimensions follows readily from the analysis presented below, and
the same conclusions apply.) The new analysis recovers the salient
features of the instability described in \citep{VayJCP2011}, including
the existence of the special time step. Growth rates are calculated
for various cases of ultra-relativistic drifting plasmas and shown
to match closely the growth rates obtained using the PIC code WARP
\citep{Warp}. Additionally, an alternative field interpolation algorithm
is proposed for which instabilities are almost completely eliminated
for a particular time step. A similar type of instability was reported
in the calculation of astrophysical shocks \citep{Spitkovsky:ICNSP2011},
and the conclusions from this paper should apply readily.

A general derivation of the numerical instability dispersion relation
for multidimensional PIC codes employing the Esirkepov algorithm is
outlined in Sec. 2. The dispersion relation is specialized in Sec.
3 to a cold, relativistic beam in two dimensions for comparison with
WARP simulations. Sec. 4 provides a simple yet reasonably accurate
analytical expression for maximum numerical instability growth rates
and, thereby, identifies time steps for which growth rates are significantly
reduced, or even eliminated under some conditions. Then, the dispersion
relation is solved numerically for a range of parameters and compared
with WARP results in Sec. 5. (Most of these analytical and numerical
calculations were performed using Mathematica \citep{Mathematica8}).
Finally, Sec. 6 presents WARP simulations, demonstrating the near
absence of numerical instabilities for an appropriately chosen field
interpolation scheme and time step in two and three dimensions.

\section{Numerical instability dispersion relation}

The derivation here follows closely that of the general numerical
instability dispersion relation in \citep{godfrey1975canonical},
and only those steps that differ will be presented . To start, the
present derivation is based on the electromagnetic fields themselves
rather than on the potentials.
\begin{equation}
\frac{\partial\mathbf{E}}{\partial t}=\nabla\times\mathbf{B}-\mathbf{J\mathrm{,}}\label{eq:Edot}
\end{equation}
\begin{equation}
\frac{\partial\mathbf{B}}{\partial t}=-\nabla\times\mathbf{E\mathrm{.}}\label{eq:Bdot}
\end{equation}
Units are chosen such that, without loss of generality, the speed
of light and other constants are unity. If the differential equations
are replaced by corresponding finite difference equations, and the
difference equations Fourier-transformed in space and time, we obtain
expressions of the form,

\begin{equation}
[\omega]\mathbf{E=-[k]}\times\mathbf{B}+i\mathbf{J\mathrm{,}}\label{eq:Etrans}
\end{equation}

\begin{equation}
[\omega]\mathbf{B=[k]}\times\mathbf{E\mathrm{.}}\label{eq:Btrans}
\end{equation}
Brackets around quantities designate their finite difference representations.

The Esirkepov algorithm determines not the current itself but its
first derivative; see Eq. (19) of \citep{esirkepov2001exact}. The
Fourier transform of this equation can be written as, 

\begin{equation}
\left\{ \begin{array}{c}
W_{x}\\
W_{y}\\
W_{z}
\end{array}\right\} =-i\Delta t\left\{ \begin{array}{c}
[k_{x}]\mathscr{\mathcal{J}}_{x}\\
{}[k_{y}]\mathcal{J}_{y}\\
{}[k_{z}]\mathcal{J}_{z}
\end{array}\right\} ,\label{eq:dJdx}
\end{equation}
with $\boldsymbol{\mathcal{J}}$ the current contribution of an individual
particle, and $\Delta t$ the simulation time step. $\mathbf{W}$
is further defined in terms of the current interpolation function
$S^{J}$ by Eq. (23) of \citep{esirkepov2001exact}, which when Fourier-transformed
becomes,

\begin{equation}
\left\{ \begin{array}{c}
W_{x}\\
W_{y}\\
W_{z}
\end{array}\right\} =-2iS^{J}\left\{ \begin{array}{c}
\sin\left(k_{x}^{\prime}v_{x}\frac{\Delta t}{2}\right)\left[\cos\left(k_{y}^{\prime}v_{y}\frac{\Delta t}{2}\right)\cos\left(k_{z}^{\prime}v_{z}\frac{\Delta t}{2}\right)-\frac{1}{3}\sin\left(k_{y}^{\prime}v_{y}\frac{\Delta t}{2}\right)\sin\left(k_{z}^{\prime}v_{z}\frac{\Delta t}{2}\right)\right]\\
\sin\left(k_{y}^{\prime}v_{y}\frac{\Delta t}{2}\right)\left[\cos\left(k_{z}^{\prime}v_{z}\frac{\Delta t}{2}\right)\cos\left(k_{x}^{\prime}v_{x}\frac{\Delta t}{2}\right)-\frac{1}{3}\sin\left(k_{z}^{\prime}v_{z}\frac{\Delta t}{2}\right)\sin\left(k_{x}^{\prime}v_{x}\frac{\Delta t}{2}\right)\right]\\
\sin\left(k_{z}^{\prime}v_{z}\frac{\Delta t}{2}\right)\left[\cos\left(k_{x}^{\prime}x_{y}\frac{\Delta t}{2}\right)\cos\left(k_{y}^{\prime}v_{y}\frac{\Delta t}{2}\right)-\frac{1}{3}\sin\left(k_{x}^{\prime}v_{x}\frac{\Delta t}{2}\right)\sin\left(k_{y}^{\prime}v_{y}\frac{\Delta t}{2}\right)\right]
\end{array}\right\} ,\label{eq:W}
\end{equation}
with $\mathbf{v}$ the particle velocity. Combining Eqs. (\ref{eq:dJdx})
and (\ref{eq:W}) provides the desired expression for the particle
current,

\begin{equation}
\left\{ \begin{array}{c}
\mathcal{J}_{x}\\
\mathcal{J}_{y}\\
\mathcal{J}_{z}
\end{array}\right\} =S^{J}\frac{2}{\Delta t}\left\{ \begin{array}{c}
\sin\left(k_{x}^{\prime}v_{x}\frac{\Delta t}{2}\right)\left[\cos\left(k_{y}^{\prime}v_{y}\frac{\Delta t}{2}\right)\cos\left(k_{z}^{\prime}v_{z}\frac{\Delta t}{2}\right)-\frac{1}{3}\sin\left(k_{y}^{\prime}v_{y}\frac{\Delta t}{2}\right)\sin\left(k_{z}^{\prime}v_{z}\frac{\Delta t}{2}\right)\right]/[k_{x}]\\
\sin\left(k_{y}^{\prime}v_{y}\frac{\Delta t}{2}\right)\left[\cos\left(k_{z}^{\prime}v_{z}\frac{\Delta t}{2}\right)\cos\left(k_{x}^{\prime}v_{x}\frac{\Delta t}{2}\right)-\frac{1}{3}\sin\left(k_{z}^{\prime}v_{z}\frac{\Delta t}{2}\right)\sin\left(k_{x}^{\prime}v_{x}\frac{\Delta t}{2}\right)\right]/[k_{y}]\\
\sin\left(k_{z}^{\prime}v_{z}\frac{\Delta t}{2}\right)\left[\cos\left(k_{x}^{\prime}x_{y}\frac{\Delta t}{2}\right)\cos\left(k_{y}^{\prime}v_{y}\frac{\Delta t}{2}\right)-\frac{1}{3}\sin\left(k_{x}^{\prime}v_{x}\frac{\Delta t}{2}\right)\sin\left(k_{y}^{\prime}v_{y}\frac{\Delta t}{2}\right)\right]/[k_{z}]
\end{array}\right\} .\label{eq:Jpart}
\end{equation}
This expression must, of course, be integrated over the linearized
particle distribution function to obtain the total current. Note that
Eq. (\ref{eq:Jpart}) reduces to $\boldsymbol{\mathcal{J}}=\mathit{\mathbf{v}}$
in the limit of vanishing time step and cell size, as it should.

Modeling relativistic simulations requires replacing $d\mathbf{v}/dt$
by $d\mathbf{p}/dt$, with $\mathbf{p=\gamma\mathbf{v}}$ the relativistic
momentum and $\gamma$ the relativistic energy, in Eq. (16) of \citep{godfrey1975canonical}.
This change flows through to Eqs. (17) and (18), in which 
\begin{equation}
\frac{\partial}{\partial\mathbf{p}}=\frac{1}{\gamma}\frac{\partial}{\partial\mathbf{v}}-\frac{\mathbf{p}}{\gamma^{3}}\mathbf{p\cdot\frac{\partial}{\partial\mathbf{v}}}\label{eq:ddp}
\end{equation}
replaces $\partial/\partial\mathbf{v}$. The force in Eqs. (17) -
(19) of \citep{godfrey1975canonical} applied to the particles is
$\mathbf{E}+\mathbf{v}$$\times\mathbf{B}$, with the components of
$\mathbf{E}$ and $\mathbf{B}$ multiplied by the Fourier transforms
of their respective interpolations functions:
\begin{equation}
\left\{ \begin{array}{c}
F_{x}\\
F_{y}\\
F_{z}
\end{array}\right\} =\left\{ \begin{array}{c}
S^{E_{x}}E_{x}+v_{y}S^{B_{z}}B_{z}-v_{z}S^{B_{y}}B_{y}\\
S^{E_{y}}E_{y}+v_{z}S^{B_{x}}B_{x}-v_{x}S^{B_{z}}B_{z}\\
S^{E_{z}}E_{z}+v_{x}S^{B_{y}}B_{y}-v_{y}S^{B_{x}}B_{x}
\end{array}\right\} .\label{eq:F}
\end{equation}
In contrast to \citep{godfrey1975canonical}, the Fourier-transformed
field interpolation functions are not assumed to be identical.

Replacing appropriate parts of Eqs. (18) and (23) of \citep{godfrey1975canonical}
by the corresponding terms from Eqs. (\ref{eq:Jpart}) - (\ref{eq:F})
yields
\begin{equation}
\mathbf{J}=\sum_{m}\int\mathbf{F\cdot\frac{\partial}{\partial\mathbf{p}}\,\mathcal{\boldsymbol{J}}\,\csc}\left[\left(\omega-\mathbf{k^{\prime}\cdot v}\right)\frac{\Delta t}{2}\right]\frac{\Delta t}{2}f\,\mathrm{d}^{3}\mathbf{v}\label{eq:J}
\end{equation}
summed over spatial aliases $m_{z}$ and $m_{x}$, as defined in \citep{godfrey1975canonical}.
The determinant of the 6x6 matrix comprised of Eqs. (\ref{eq:Etrans}),
(\ref{eq:Btrans}), and (\ref{eq:J}) is the desired dispersion relation.
A striking difference between this and the general dispersion relation
in \citep{godfrey1975canonical} is that the present dispersion relation
contains trigonometric functions involving particle velocities.

\section{WARP 2-d dispersion relation}

For comparison with WARP two-dimensional, cold beam simulation results
\citep{VayJCP2011}, we reduce Eqs. (\ref{eq:Etrans}) and (\ref{eq:Btrans})
to a 3x3 system in $\left\{ E_{z},E_{x},B_{y}\right\} $ and perform
the integral in Eq(\ref{eq:J}) for a cold beam propagating at velocity
\textit{v} in the \textit{z}-direction. The resulting matrix equation
is
\begin{equation}
\left(\begin{array}{ccc}
\xi_{z,z}+[\omega] & \xi_{z,x} & \xi_{z,y}+[k_{x}]\\
0 & \xi_{x,x}+[\omega] & \xi_{x,y}-[k_{z}]\\
D_{x}^{*}[k_{x}] & -D_{z}^{*}[k_{z}] & [\omega]
\end{array}\right)\left(\begin{array}{c}
E_{z}\\
E_{x}\\
B_{y}
\end{array}\right)=0.\label{eq:M3x3}
\end{equation}
$D_{z}^{*}$ and $D_{x}^{*}$ are introduced at this point to accommodate
the Cole-Karkkainnen field solver, sometimes used in WARP; it is discussed
near the end of this section. The quantities $\xi$ are employed purely
for notational simplicity.
\begin{equation}
\xi_{z,z}\equiv-n\gamma^{-2}\sum_{m}S^{J}S^{E_{z}}\csc^{2}\left[\left(\omega-k_{z}^{\prime}v\right)\frac{\Delta t}{2}\right]\frac{\Delta t}{2}\sin\left(\omega\frac{\Delta t}{2}\right)k_{z}^{\prime}/[k_{z}]\mathrm{,}\label{eq:Mzz}
\end{equation}
\begin{equation}
\xi_{z,x}\equiv-n\sum_{m}S^{J}S^{E_{x}}\csc\left[\left(\omega-k_{z}^{\prime}v\right)\frac{\Delta t}{2}\right]\cot\left[\left(\omega-k_{z}^{\prime}v\right)\frac{\Delta t}{2}\right]\frac{\Delta t}{2}\sin\left(k_{z}^{\prime}v\frac{\Delta t}{2}\right)k_{x}^{\prime}/[k_{z}],\label{eq:Mzx}
\end{equation}
\begin{equation}
\xi_{z,y}\equiv nv\sum_{m}S^{J}S^{B_{y}}\csc\left[\left(\omega-k_{z}^{\prime}v\right)\frac{\Delta t}{2}\right]\cot\left[\left(\omega-k_{z}^{\prime}v\right)\frac{\Delta t}{2}\right]\frac{\Delta t}{2}\sin\left(k_{z}^{\prime}v\frac{\Delta t}{2}\right)k_{x}^{\prime}/[k_{z}],\label{eq:Mzy}
\end{equation}
\begin{equation}
\xi_{x,x}\equiv-n\sum_{m}S^{J}S^{E_{x}}\csc\left[\left(\omega-k_{z}^{\prime}v\right)\frac{\Delta t}{2}\right]\frac{\Delta t}{2}\cos\left(k_{z}^{\prime}v\frac{\Delta t}{2}\right)k_{x}^{\prime}/[k_{x}],\label{eq:Mxx}
\end{equation}
\begin{equation}
\xi_{x,y}\equiv nv\sum_{m}S^{J}S^{B_{y}}\csc\left[\left(\omega-k_{z}^{\prime}v\right)\frac{\Delta t}{2}\right]\frac{\Delta t}{2}\cos\left(k_{z}^{\prime}v\frac{\Delta t}{2}\right)k_{x}^{\prime}/[k_{x}].\label{eq:Mxy}
\end{equation}
summed over spatial aliases, $k_{z}^{\prime}=k_{z}+m_{z}\,2\pi/\Delta z$
and $k_{x}^{\prime}=k_{x}+m_{x}\,2\pi/\Delta x$, with $m_{z}$ and
$m_{x}$ integers. The resonances, $\omega-k_{z}^{\prime}v$, introduce
an infinity of spurious beam modes with effective charge densities
proportional to $S^{J}S^{E_{z}}$, \textit{etc}. \textit{n} is the
beam charge density divided by $\gamma$, which can be normalized
to unity. However, explicitly retaining it in the dispersion relation
sometimes is informative.

WARP employs the usual staggered spatial mesh and E-B leapfrog in
time \citep{Yee}. Hence, 
\begin{equation}
\left[\omega\right]=\sin\left(\omega\frac{\Delta t}{2}\right)/\left(\frac{\Delta t}{2}\right),\label{eq:meshom}
\end{equation}
\begin{equation}
\left[k_{z}\right]=\sin\left(k_{z}\frac{\Delta z}{2}\right)/\left(\frac{\Delta z}{2}\right),\label{eq:meshkz}
\end{equation}
\begin{equation}
\left[k_{x}\right]=\sin\left(k_{x}\frac{\Delta x}{2}\right)/\left(\frac{\Delta x}{2}\right).\label{eq:meshkx}
\end{equation}
Also as usual, WARP employs splines for current and field interpolation.
The Fourier transform of the current interpolation function is
\begin{equation}
S^{J}=\left[\sin\left(k_{z}^{\prime}\frac{\Delta z}{2}\right)/\left(k_{z}^{\prime}\frac{\Delta z}{2}\right)\right]^{\ell_{z}+1}\left[\sin\left(k_{x}^{\prime}\frac{\Delta x}{2}\right)/\left(k_{x}^{\prime}\frac{\Delta x}{2}\right)\right]^{\ell_{x}+1},\label{eq:SJ}
\end{equation}
$\ell_{z}$ and $\ell_{x}$ are the orders of the current interpolation
splines in the \textit{z}- and \textit{x}-directions. So, for instance,
an exponent of 2 in Eq. (\ref{eq:SJ}) corresponds to linear interpolation,
and of 4 to cubic interpolation. Analogous definitions apply to the
three field interpolation functions, but the spline orders need not
be the same. WARP typically employs field interpolation splines like
those of the currents but with the $E_{z}$ splines one order lower
in \textit{z}, the $E_{x}$ splines one order lower in \textit{x},
and the $B_{y}$ splines one order lower in both. (This particular
choice of spline orders is derivable by Galerkin's method \citep{godfrey1975galerkin}
and has superior energy conservation properties \citep{lewis1972variational,langdon1973energy,BirdsallLangdon}.
It will be referred to subsequently as ``Galerkin field interpolation''.)
\begin{equation}
S^{E_{z}}=\left[\sin\left(k_{z}^{\prime}\frac{\Delta z}{2}\right)/\left(k_{z}^{\prime}\frac{\Delta z}{2}\right)\right]^{\ell_{z}}\left[\sin\left(k_{x}^{\prime}\frac{\Delta x}{2}\right)/\left(k_{x}^{\prime}\frac{\Delta x}{2}\right)\right]^{\ell_{x}+1}\left(-1\right)^{m_{z}},\label{eq:SEzG}
\end{equation}
\begin{equation}
S^{E_{x}}=\left[\sin\left(k_{z}^{\prime}\frac{\Delta z}{2}\right)/\left(k_{z}^{\prime}\frac{\Delta z}{2}\right)\right]^{\ell_{z}+1}\left[\sin\left(k_{x}^{\prime}\frac{\Delta x}{2}\right)/\left(k_{x}^{\prime}\frac{\Delta x}{2}\right)\right]^{\ell_{x}}\left(-1\right)^{m_{x}},\label{eq:SExG}
\end{equation}
\begin{equation}
S^{B_{y}}=\cos\left(\omega\frac{\Delta t}{2}\right)\left[\sin\left(k_{z}^{\prime}\frac{\Delta z}{2}\right)/\left(k_{z}^{\prime}\frac{\Delta z}{2}\right)\right]^{\ell_{z}}\left[\sin\left(k_{x}^{\prime}\frac{\Delta x}{2}\right)/\left(k_{x}^{\prime}\frac{\Delta x}{2}\right)\right]^{\ell_{x}}\left(-1\right)^{m_{z}+m_{x}}.\label{eq:SByG}
\end{equation}
The alias phase factors appearing at the ends of Eqs. (\ref{eq:SEzG})
- (\ref{eq:SByG}) arise from the half-cell offsets from the current
interpolation mesh of the corresponding fields. Averaging $B_{y}$
in time before applying it to particles causes the factor $\cos\left(\omega\frac{\Delta t}{2}\right)$
in Eq.(\ref{eq:SByG}).

Another credible choice of field interpolation functions is splines
of the same order as those for the current interpolation function,
in which case Eqs. (\ref{eq:SEzG}) - (\ref{eq:SByG}) contain only
powers of $\ell_{x}+1$ and $\ell_{z}+1$. The powers of -1 are unchanged.
This seemingly minor change has a significant impact on numerical
stability for some choices of $\Delta t$. (It will be referred to
subsequently as ``uniform field interpolation''.)

The Cole-Karkkainnen field solver \citep{ColeIEEE1997,ColeIEEE2002,KarkICAP06},
mentioned above, increases the Courant limit on the simulation time
step and in some cases reduces numerical dispersion in the electromagnetic
fields. It is discussed in some detail in Sec. 2.2 of \citep{VayJCP2011}.
For our purposes,
\begin{equation}
D_{z}^{*}=1-4\beta_{x}\sin^{2}\left(k_{x}\frac{\Delta x}{2}\right),\label{eq:Dz}
\end{equation}
\begin{equation}
D_{x}^{*}=1-4\beta_{z}\sin^{2}\left(k_{z}\frac{\Delta z}{2}\right).\label{eq:Dx}
\end{equation}
For $\Delta x=\Delta z$, the choice $\beta_{x}=\beta_{z}=1/8$ relaxes
the Courant limit to $\Delta t<\Delta z$, while minimizing numerical
dispersion in the vacuum fields along major axes.

Finally, we note that $m_{x}$ alias terms in the dispersion relation
can be summed explicitly by means of Eqs. (1.421.3) and (1.422.3)
of \citep{GradshteynRyzhik} or derivatives thereof, once choices
have been made for the interpolation functions. For example, the $k_{x}^{\prime}$-dependent
terms in $\xi_{z,z}$ sum to
\begin{equation}
\sum_{m_{x}}\left[\sin\left(k_{x}^{\prime}\frac{\Delta x}{2}\right)/\left(k_{x}^{\prime}\frac{\Delta x}{2}\right)\right]^{4}=\left[2\cos\left(k_{x}\frac{\Delta x}{2}\right)+1\right]/3\label{eq:Sumzz}
\end{equation}
for $\ell_{x}=1$. Note that the $m_{x}=0$ term alone has the value
$\left(2/\pi\right)^{4}$ for $k_{x}$ near its maximum value, $\pi/\Delta x$.
In contrast, the sum has the value $\nicefrac{1}{3}$ there. (Most
of the difference is due to the $m_{x}=-1$ alias, which is typical.)
Since, as we shall see, peak growth rates typically scale as the cube
root of such sums, the difference in predicted peak growth rates is
of order 20\%.

\section{Approximate peak growth rates}

$\xi_{z,z}$, defined in Eq. (\ref{eq:Mzz}), scales as $\gamma^{-2}$
(with \textit{n} held constant) and can be ignored for highly relativistic
calculations, on which this paper focuses. Likewise, $1-v\simeq\gamma^{-2}/2$,
and can be set to zero. Additionally,
\begin{equation}
\xi_{z,x}\xi_{x,y}-\xi_{z,y}\xi_{z,y}=0\label{eq:minor}
\end{equation}
 is satisfied for individual modes and is satisfied approximately
for cross-products between modes. With these assumptions the dispersion
relation (the determinant of Eq. (\ref{eq:M3x3})) has the form
\begin{equation}
C_{0}+n\sum_{m_{z}}C_{1}\csc\left[\left(\omega-k_{z}^{\prime}\right)\frac{\Delta t}{2}\right]+n\sum_{m_{z}}C_{2}\csc^{2}\left[\left(\omega-k_{z}^{\prime}\right)\frac{\Delta t}{2}\right]=0.\label{eq:drform}
\end{equation}
with $C_{0}$ the vacuum dispersion function, 
\begin{equation}
C_{0}=\left[\omega\right]^{2}-D_{z}^{*}\left[k_{x}\right]^{2}-D_{x}^{*}\left[k_{z}\right]^{2},\label{eq:C0}
\end{equation}
and
\begin{equation}
C_{1}=-\frac{C_{0}}{\left[\omega\right]}\frac{\Delta t}{2}\sum_{m_{x}}\frac{k_{x}^{\prime}}{\left[k_{x}\right]}S^{J}S^{E_{x}}\cos\left(k_{z}^{\prime}\frac{\Delta t}{2}\right)-D_{z}^{*}\frac{[k_{z}]^{2}}{\left[k_{x}\right]}\frac{\Delta t}{2}\sum_{m_{x}}k_{x}^{\prime}S^{J}\left(\frac{S^{E_{x}}}{\left[\omega\right]}-\frac{S^{B_{y}}}{[k_{z}]}\right)\cos\left(k_{z}^{\prime}\frac{\Delta t}{2}\right),\label{eq:C1}
\end{equation}
\begin{equation}
C_{2}=D_{x}^{*}[k_{x}]\frac{\Delta t}{2}\sum_{m_{x}}k_{x}^{\prime}S^{J}\left(\frac{S^{E_{x}}}{\left[\omega\right]}-\frac{S^{B_{y}}}{[k_{z}]}\right)\cos\left[\left(\omega-k_{z}^{\prime}\right)\frac{\Delta t}{2}\right]\sin\left(k_{z}^{\prime}\frac{\Delta t}{2}\right).\label{eq:C2}
\end{equation}
Eq.(\ref{eq:drform}) reduces, of course, to $C_{0}+n=0$ in the limit
of vanishing time step and cell size. All the beam modes in Eq.(\ref{eq:drform})
are numerical artifacts, even the $m_{z}=0$ mode.

\begin{figure}
\begin{centering}
\includegraphics{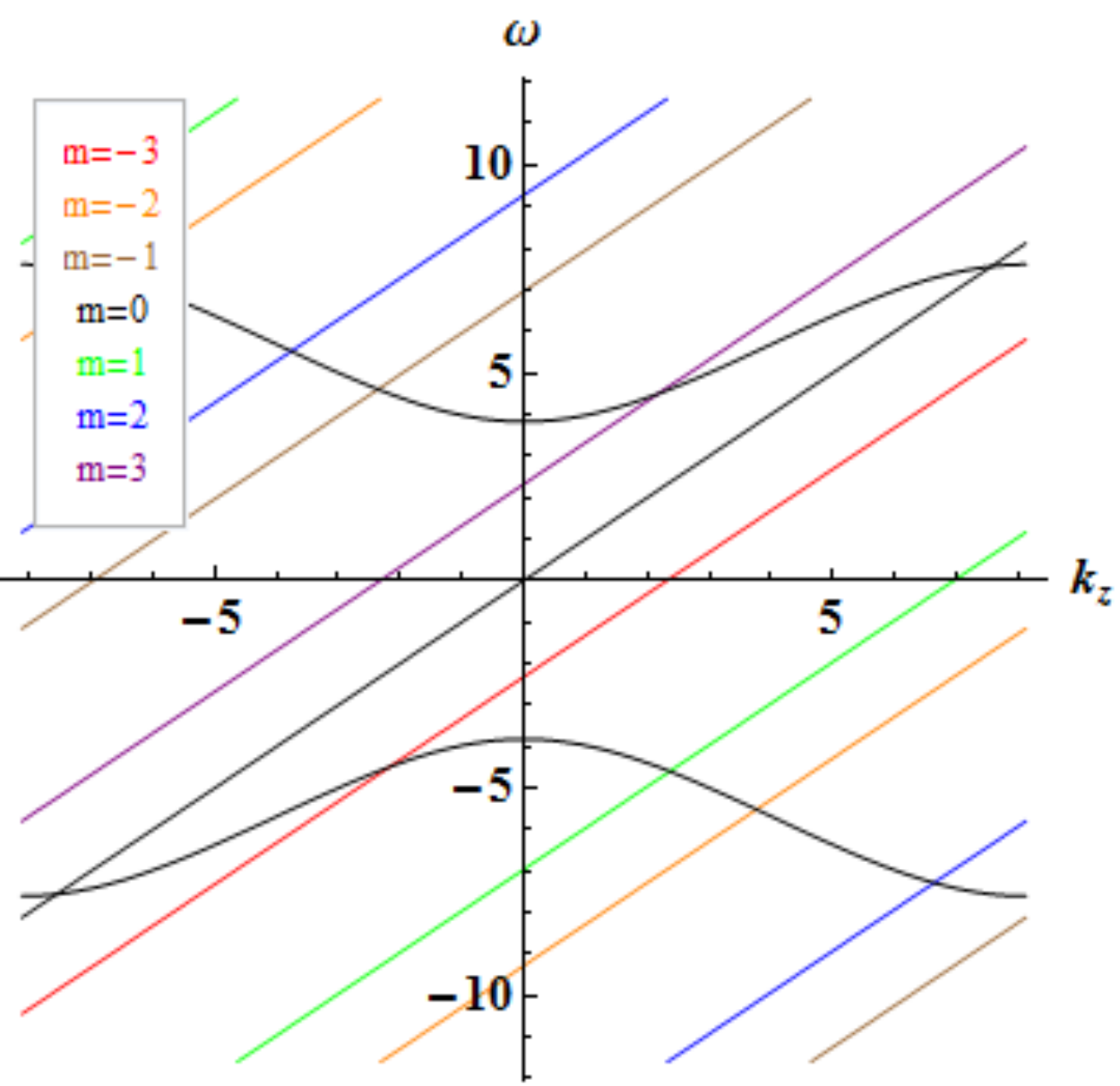}
\par\end{centering}

\caption{\label{fig:Normal-mode-diagram}Normal mode diagram for$\frac{\Delta t}{\Delta z}=0.7$,
$\beta=0$, and $k_{x}=\nicefrac{1}{2}\,\frac{\pi}{\Delta x}$, showing
numerically distorted electromagnetic modes and spurious beam modes,
$m_{z}=\left[-3,\,3\right]$. Numerical Cherenkov instabilities occur
near mode intersections.}
\end{figure}

Coupling between these beam numerical modes and electromagnetic modes
(the roots of $C_{0}=0$) gives rise to what has become known as the
numerical Cherenkov instability \citep{godfrey1974numerical,godfrey1979electro},
which can be quite virulent. Fig. \ref{fig:Normal-mode-diagram} is
a typical normal mode diagram, showing the two electromagnetic modes
and beam aliases $m_{z}=$ {[}-3, 3{]} for $\Delta t=0.7\Delta z$,
$\beta_{x}=\beta_{z}=0$, and $k_{x}=\nicefrac{1}{2}\,\frac{\pi}{\Delta x}$.
(Unless otherwise noted, other parameters for this and other figures
are $n=1$ and $\Delta x=\Delta z=0.3868$.) Fig. \ref{fig:Resonance curves}
depicts the locations in \textit{k}-space of normal mode intersections,
such as those in Fig. \ref{fig:Normal-mode-diagram}, as $k_{x}$
is varied. 

Comparing Fig. \ref{fig:Resonance curves} with corresponding WARP
results in Fig. \ref{fig:Ez sample} indicates that the strongest
instabilities lie along the $m_{z}$= -1 and 0 resonance curves at
larger $k_{x}$. (The WARP simulations were performed on a $128\times128$
square grid with periodic boundary conditions and a uniformly distributed
plasma moving axially at an energy of $\gamma=130$, seeded with a
small random transverse velocity. Plots similar to Fig. \ref{fig:Ez sample}
appear in \citep{UCLA2012AAC,Vay2012AAC}.) Also visible, although
just barely, are much more slowly growing instabilities along the
$m_{z}$= +1 and $m_{z}$= -2 resonance curves. We now proceed to
estimate these instability growth rates.

Resonance curves, such as those in Fig. \ref{fig:Resonance curves},
are given by Eq. (\ref{eq:C0}) with $\omega$ replaced by $k_{z}^{\prime}$
, solved for $k_{x}$ as a function of $k_{z}^{\prime}$. (Recall
that $\sin^{2}\left(k_{z}^{\prime}\frac{\Delta z}{2}\right)=\sin^{2}\left(k_{z}\frac{\Delta z}{2}\right)$.)
\begin{equation}
k_{x}^{r}=\frac{2}{\Delta x}\arcsin\left(\sqrt{\frac{\left(\frac{\Delta t}{\Delta z}\right)^{2}\sin^{2}\left(k_{z}^{\prime}\frac{\Delta t}{2}\right)-\left(\frac{\Delta x}{\Delta z}\right)^{2}\sin^{2}\left(k_{z}^{\prime}\frac{\Delta z}{2}\right)}{1-4\sin^{2}\left(k_{z}^{\prime}\frac{\Delta z}{2}\right)\left(\beta_{x}+\beta_{z}\left(\frac{\Delta x}{\Delta z}\right)^{2}\right)}}\right)\label{eq:kx resonance}
\end{equation}
To obtain an estimate of the numerical instability growth rate along
a resonance curve, we expand $C_{0}$ and the cosecants in Eq. (\ref{eq:drform})
to first order in $\left(\omega-k_{z}^{\prime}\right)$, set $C_{1}=0$,
and set $\omega=k_{z}^{\prime}$ in $C_{2}$. The resulting cubic
equation has one unstable root,
\begin{equation}
Im\left(\omega\right)\simeq\frac{\sqrt{3}}{2}\sqrt[3]{\frac{n}{2}D_{x}^{*}[k_{x}]\sum_{m_{x}}k_{x}^{\prime}S^{J}\left|\frac{\frac{\Delta t}{2}S^{E_{x}}}{\sin\left(k_{z}^{\prime}\frac{\Delta t}{2}\right)}-\frac{\frac{\Delta z}{2}S^{B_{y}}}{\sin\left(k_{z}^{\prime}\frac{\Delta z}{2}\right)}\right|\csc\left(k_{z}^{\prime}\frac{\Delta t}{2}\right)},\label{eq:cubic growth}
\end{equation}
evaluated at $k_{x}=k_{x}^{r}$. Although it may appear that Eq. (\ref{eq:cubic growth})
becomes singular when $k_{z}^{\prime}$ approaches zero, $k_{x}^{r}$
approaches zero there also, as $k_{z}^{\prime}$$^{2}$. Consequently,
the growth rate vanishes in that limit.

For completeness, we note that instability also occurs off-resonance
when $C_{0}C_{2}>C_{1}^{\,2}/4$, evaluated at $\omega\simeq k_{z}^{\prime}$
and arbitrary $k_{x}$. The resulting growth rate is
\begin{equation}
Im\left(\omega\right)\simeq\frac{\sqrt{C_{1}^{\,2}/4-C_{0}C_{2}}}{C_{0}}.\label{eq:quadratic growth}
\end{equation}
Although off-resonance growth is weaker than on-resonance, it often
occurs at smaller $k_{z}$, where it may be more difficult to filter.
(The residual instabilities after digital filtering discussed in the
fourth paragraph of Sec. 5 are, for instance, of this sort.)

Fig. \ref{fig:Galerkin approx growth} displays maximum instability
growth rates for the Galerkin field interpolation algorithm as $\Delta t/\Delta z$
varies over its range of allowed values for $\beta_{z}=\beta_{x}$
(collectively, $\beta)$ = 0 and $\nicefrac{1}{8}$. (Intermediate
values of $\beta$ produce curves intermediate in shape.) The pronounced
dip in both curves, at $\Delta t/\Delta z$$\approx0.66$ for $\beta=0$
and 0.69 for $\beta=\nicefrac{1}{8}$, previously has been observed
in simulations \citep{VayJCP2011}. It occurs because $Im\left(\omega\right)$
vanishes for some value of $k_{z}$, which occurs when
\begin{equation}
\frac{\Delta t}{\Delta z}\sin^{2}\left(k_{z}^{\prime}\frac{\Delta z}{2}\right)=k_{z}^{\prime}\frac{\Delta z}{4}\sin\left(k_{z}^{\prime}\Delta t\right).\label{eq:Galerkin zeropoint}
\end{equation}
Eq. (\ref{eq:Galerkin zeropoint}) has solutions for the $m_{z}$=
-1 and 0 resonances only over a narrow range of time steps, $\sqrt{2}/2\geq\Delta t/\Delta z\gtrsim0.65$.
Precisely where the minimum falls within this range depends on algorithmic
details.

Similarly, Fig. \ref{fig:Uniform approx growth} displays maximum
instability growth rates for the uniform field interpolation algorithm
as $\Delta t/\Delta z$ varies over its range of allowed values for
$\beta$ = 0 and $\nicefrac{1}{8}$. For all values of $\beta$, the
growth rate vanishes at $\Delta t/\Delta z$ = $\nicefrac{1}{2}$.
Why this should be so is evident from 
\begin{equation}
\frac{\Delta t}{\Delta z}\sin\left(k_{z}^{\prime}\frac{\Delta z}{2}\right)=\frac{1}{2}\sin\left(k_{z}^{\prime}\Delta t\right),\label{eq:Uniform zeropoint}
\end{equation}
which differs from its Galerkin counterpart, Eq. (\ref{eq:Galerkin zeropoint}),
by a factor of $\sin\left(k_{z}^{\prime}\frac{\Delta z}{2}\right)/\left(k_{z}^{\prime}\frac{\Delta z}{2}\right)$.
Eq. (\ref{eq:Uniform zeropoint}) is satisfied for all $k_{z}^{\prime}$
at $\Delta t/\Delta z$ = $\nicefrac{1}{2}$, and for no values (apart
from 0) of $k_{z}^{\prime}$ otherwise. 

Eq. (\ref{eq:cubic growth}) also provides a simple means for estimating
the effect of current filtering on numerical Cherenkov instabilities,
because the Fourier transform of the digital filtering function appears
simply as a factor multiplying \emph{n}. Given the substantial growth
rates of this instability, filtering must reduce currents in regions
of \emph{k}-space where the instability is strong by some three orders
of magnitude. Of course, any physical phenomena occurring in those
same regions also will be suppressed. Using higher order interpolation
(\emph{i.e.}, larger $\ell$'s in Eqs. (\ref{eq:SJ}) - (\ref{eq:SByG}))
also reduces numerical instability growth, especially for higher order
aliases. However, for typical simulation parameters it reduces the
$m_{z}$= -1 and 0 instability growth rates by comparable, modest
factors. Employing cubic rather than linear splines, for instance,
would be expected to reduce maximum growth rates by of order $\left(2/\pi\right)^{\nicefrac{4}{3}}$.
On this basis current digital filtering usually is more cost effective
than higher order interpolation for suppressing numerical Cherenkov
instabilities.

\section{Numerical solutions}

Reliably finding the roots of Eq. (\ref{eq:M3x3}) can be accomplished
as follows. Given how strongly even linear interpolation suppresses
all but the first few aliases, we safely can truncate the infinite
series in $m_{z}$ to a range of, say, {[}-3, 3{]}. (Indeed, the smaller
range {[}-1, 0{]} works fairly well in most cases.) Then, if the aliases
are well separated in $\omega-k$ space (as they are in, for instance,
Fig. \ref{fig:Normal-mode-diagram}), the growth rates for any particular
alias can be evaluated with reasonable accuracy by expanding the dispersion
relation as a fourth-order power series in $\left(\omega-k_{z}^{\prime}v\right)$
for the $k_{z}^{\prime}$ in question and calculating all roots with
a polynomial root finder. On the other hand, if aliases are separated
in frequency by only a few times the typical growth rates, the expansion
converges slowly, and an iterative solution is required. The Mathematica
\citep{Mathematica8} FindRoot routine was used for the results that
follow in this section, with three real roots or one real root and
one conjugate pair of roots found per alias. Evaluations were performed
on a 65x65 array in \emph{k}-space, consistent with the 128x128 spacial
grid used in WARP for comparable simulations. Obtaining results for
a typical set of parameters required about 15 minutes on a 2.8 GHz,
2 processor desktop computer. 

Fig. \ref{fig:Numerical growth sample} presents numerical growth
rate predictions corresponding to the WARP results in Fig. \ref{fig:Ez sample}.
The $m_{z}$= -1 alias dominates the growth spectrum with a maximum
growth rate of 0.56 at short wavelengths in x. (The approximate growth
rate based on the analysis in the previous section is 0.48.) Also
visible is the fast growing $m_{z}$= 0 alias. The much weaker $m_{z}$=
-2 and +1 aliases are evident at smaller $k_{z}$. As noted in the
previous section, the $m_{z}$= -1, 0, and +1 aliases all can be seen
in Fig. \ref{fig:Ez sample}, although the last of these aliases is
faint, consistent with its relatively slow growth. Fig. \ref{fig:Growth rate sample}
depicts grow rates measured in this WARP simulation (actually the
average of one hundred such simulations). The agreement between Figs.
\ref{fig:Numerical growth sample} and \ref{fig:Growth rate sample}
is very good, especially when one considers the difficulty in measuring
smaller growth rates in simulations, where nonlinear mode coupling
and thermal noise can be significant. Thus, the method used to determine
automatically the growth rates in WARP works well for the largest
growth rates, which are of most interest in any particular simulation,
but not so well for the smallest growth rates.

Maximum numerical growth rates observed in WARP for the Galerkin and
uniform current interpolation algorithms with $\beta$=0 and $\nicefrac{1}{8}$
are compared with the predictions of linear theory in Figs. \ref{fig:Galerkin growth scan}
and \ref{fig:Uniform growth scan}. Agreement between theory and simulation
is very good. Qualitative agreement with the analytical estimates
of the previous section is quite acceptable. The sudden rise of growth
when $\Delta t/\Delta z$ nears unity for $\beta=\nicefrac{1}{8}$
comes from an instability of the field solver algorithm at the Nyquist
limit and is mitigated by using one or more passes of bilinear filtering
of the current density, as explained in Appendix A of \citep{VayJCP2011}
and shown below.

Fig. \ref{fig:Smoothing} illustrates the effects of digital filtering
and of higher order interpolation, in this case ten passes of the
bilinear filter (including two compensation steps) described in \citep{VayJCP2011},
cubic or linear interpolation in \emph{z} with Galerkin field gathering,
and $\beta=\nicefrac{1}{8}$. The digital filter has the effect of
multiplying \emph{n} in the dispersion relation by
\begin{equation}
\cos^{16}\left(k_{z}\frac{\Delta z}{2}\right)\left(5-4\cos^{2}\left(k_{z}\frac{\Delta z}{2}\right)\right)^{2}\cos^{16}\left(k_{x}\frac{\Delta x}{2}\right)\left(5-4\cos^{2}\left(k_{x}\frac{\Delta x}{2}\right)\right)^{2}.
\end{equation}
It effectively eliminates numerical instabilities for $k_{z}\Delta z/\pi\gtrsim0.2$
or $k_{x}\Delta x/\pi\gtrsim0.2$. With linear interpolation the $m_{z}=0$
alias dominates the numerical instabilty growth rate except in the
vicinity of $\Delta t/\Delta z\approx0.69$, where the $m_{z}=+1$
alias dominates. Growth rates are reduced by roughly a factor of four
compared to those in Fig. \ref{fig:Galerkin growth scan}. Cubic interpolation
has negligible effect on the $m_{z}=0$ alias but almost completely
suppresses the $m_{z}=+1$ alias. The minimum growth rate, now at
$\Delta t/\Delta z\approx0.70$, drops by a further factor of three.
(Measuring the WARP instability growth rates for Fig. \ref{fig:Smoothing}
was particularly challenging due to competition between the weak numerical
instabilities, and thermal and nonlinear effects.)

As a further comparison between linear theory, Fig. \ref{fig:DISP multi-alias},
and WARP results, Fig.\ref{fig:WARP multi-alias} (also averaged over
100 simulations), we present growth rates for Galerkin current interpolation
with $\frac{\Delta t}{\Delta z}=0.69$, and $\beta=\nicefrac{1}{8}$.
The dominant alias is $m_{z}=+1$, occurring at the rather small axial
wave numbers, $1.5<k_{z}<3.5$, and at most $k_{x}$ values away from
the $k_{z}$-axis. Modestly to the right is the $m_{z}=-2$ alias,
occurring at $3<k_{z}<4$ for large values of $k_{x}$. Generally,
we expect the $m_{z}=+1$ and -2 aliases to have comparable growth
rates, just as the $m_{z}=0$ and -1 aliases typically do. Finally,
the $m_{z}=-3$ alias is modestly above background on the far right.
For all these modes, theory and simulation growth rates agree to within
about 15\%. However, a region of reduced growth rate in the band $5<k_{z}<6$
occurs only in Fig. \ref{fig:WARP multi-alias}, although it can be
produced in Fig. \ref{fig:DISP multi-alias} by artificially removing
the off-resonance $m_{z}=-1$ contribution. This minor discrepancy
is apparent only for parameters very near those listed in this paragraph.

\section{Application to the modeling of laser plasma acceleration}

As a verification that the theory that has been developed in this
paper applies to the modeling of LPAs, series of two and three dimensional
simulations of a 100 MeV class LPA stage were performed, focusing
on the plasma wake formation, using the parameters given in table
\ref{TableLPA}. The velocity of the wake in the plasma corresponds
to $\gamma\simeq13.2$, and the simulations were performed in a boosted
frame of $\gamma_{f}=13.$ 

Reference simulations were run in two and three dimensions for conditions
where no instability developed, and the final total field energy $W_{f0}$
was recorded as a reference value in each case. Runs then were conducted
for the Yee ($\beta=0$) and Cole-Karkkainnen ($\beta=1/8$) solvers,
with Galerkin and uniform field interpolations. The final energy $W_{f}$
was recorded and divided by the reference energy $W_{f0}$. The ratio
$W_{f}/W_{f0}$ is plotted versus time step from two dimensional simulations
in Fig. \ref{fig:WARP saturation} and from three dimensional simulations
in Fig. \ref{fig:WARP saturation-3D}, using linear current deposition
and no smoothing of current and fields. Following the theoretical
predictions, for the Galerkin interpolation scheme the instability
is minimal around $\triangle t/\triangle z\approx0.65$ when $\beta=0$
and around $\triangle t/\triangle z\approx0.69$ when $\beta=1/8$,
while for the uniform interpolation scheme the instability is minimal
around $\triangle t/\triangle z\approx0.5$. The ratio $W_{f}/W_{f0}$
also is plotted versus time step from two dimensional simulations
in Fig. \ref{fig:WARP saturation-2D-smooth} and from three dimensional
simulations in Fig. \ref{fig:WARP saturation-3D-smooth}, using, as
is common practice in the modeling of laser plasma stages, cubic current
deposition and 1 pass of bilinear smoothing plus compensation of current
and fields gathered onto macroparticles. The beneficial impact on
stability of smoothing and high order deposition is evident from the
relatively wide band of stability that is available around $\triangle t/\triangle z\approx0.5$
with uniform gather, and the narrower band of stability that is available
around $\triangle t/\triangle z\approx0.7$ with Galerkin gather.
This verifies that the theoretical results apply to real case simulations
in two and three dimensions.

\begin{table}[htd]
\caption{List of parameters for a LPA stage simulation at 100 MeV}

\begin{centering}
\begin{tabular}{lcc}
\hline 
plasma density on axis  & $n_{e}$  & $10^{19}$~cm$^{-3}$\tabularnewline
plasma longitudinal profile  &  & flat\tabularnewline
plasma length  & $L_{p}$  & $1.5$ mm\tabularnewline
plasma entrance ramp profile  &  & half sine\tabularnewline
plasma entrance ramp length  &  & $20$ $\mu$m\tabularnewline
\hline 
laser profile  &  & $a_{0}\exp\left(-r^{2}/2\sigma^{2}\right)\sin\left(\pi z/3L\right)$\tabularnewline
normalized vector potential  & $a_{0}$  & $1$\tabularnewline
laser wavelength  & $\lambda$  & $0.8$ $\mu$m\tabularnewline
laser spot size (RMS)  & $\sigma$  & $8.91$ $\mu$m\tabularnewline
laser length (HWHM)  & $L$  & $3.36$ $\mu$m\tabularnewline
normalized laser spot size  & $k_{p}\sigma$  & $5.3$\tabularnewline
normalized laser length  & $k_{p}L$  & $2$\tabularnewline
\hline 
cell size in x  & $\Delta x$  & $\lambda/32$\tabularnewline
cell size in y (3D only) & $\Delta y$  & $\lambda/32$\tabularnewline
cell size in z  & $\Delta z$  & $\lambda/32$\tabularnewline
\# of plasma particles/cell  &  & 1 macro-e$^{-}$+1 macro-p$^{+}$\tabularnewline
\hline 
\end{tabular}
\par\end{centering}

\label{TableLPA} 
\end{table}

\section{Conclusion}

The numerical stability properties of multidimensional PIC codes employing
the Esirkepov current algorithm have been derived. Just as in PIC
codes employing earlier current algorithms, here also fast-growing
numerical instabilities are predicted for relativistic beam simulations.
These instabilities can, of course, be reduced significantly by short
wavelength digital filtering. However, time steps have been identified
at which instability growth is reduced even without filtering. Particularly
noteworthy is uniform field interpolation with $\Delta t/\Delta z$
= $\nicefrac{1}{2}$ and any value of $\beta$, for which simulations
are numerically stable in the large $\gamma$ limit. These results
have been confirmed with the WARP simulation code.

Additionally, WARP LPA simulations performed using uniform field interpolation
with $\Delta t/\Delta z$ = $\nicefrac{1}{2}$ have demonstrated the
practical value of this choice of parameters in two and three dimensions.
The uniform field interpolation offers much reduced growth rates,
enabling faster simulations with fewer grid cells, lower order interpolation,
and reduced digital filtering. In three dimensions, it enables existing
PIC codes that incorporate the Yee solver, but not the CK solver,
to benefit from the reduced growth rates at the special time steps
over a wider range of cell aspect ratios (for cubic cells for example,
the special time step is accessible only to the CK solver for Galerkin
gather, while it is accessible to both the Yee and the CK solvers
for uniform gather). The results that were obtained here also should
apply readily to more efficient modeling of astrophysical shocks that
use the same algorithms.

Finally, the salutary effect of trigonometric functions involving
particle velocities in the dispersion relation of the Esirkepov algorithm
suggest that further improvements in PIC code stability can be achieved
by developing field interpolation algorithms that introduce similar
trigonometric functions, perhaps along the lines of Sec. 4 in \citep{godfrey1975canonical}.

\section{Acknowledgments}

We wish to thank Irving Haber for suggesting this collaboration and
for many helpful recommendations. We also are indebted to David Grote
for support in using the code WARP at the National Energy Research
Supercomputing Center and to Andrew Moylan for advice on using Mathematica
to find arrays of roots to transcendental equations. This work was
supported in part by the Director, Office of Science, Office of High
Energy Physics, U.S. Dept. of Energy under Contract No. DE-AC02-05CH11231
and the US-DOE SciDAC ComPASS collaboration, and used resources of
the National Energy Research Scientific Computing Center.

\section*{References}

\bibliographystyle{elsarticle-num}
\bibliography{arXiv12_Godfrey_JLV}

\section*{\clearpage{}}

\begin{center}
\clearpage{}
\begin{figure}
\begin{centering}
\includegraphics{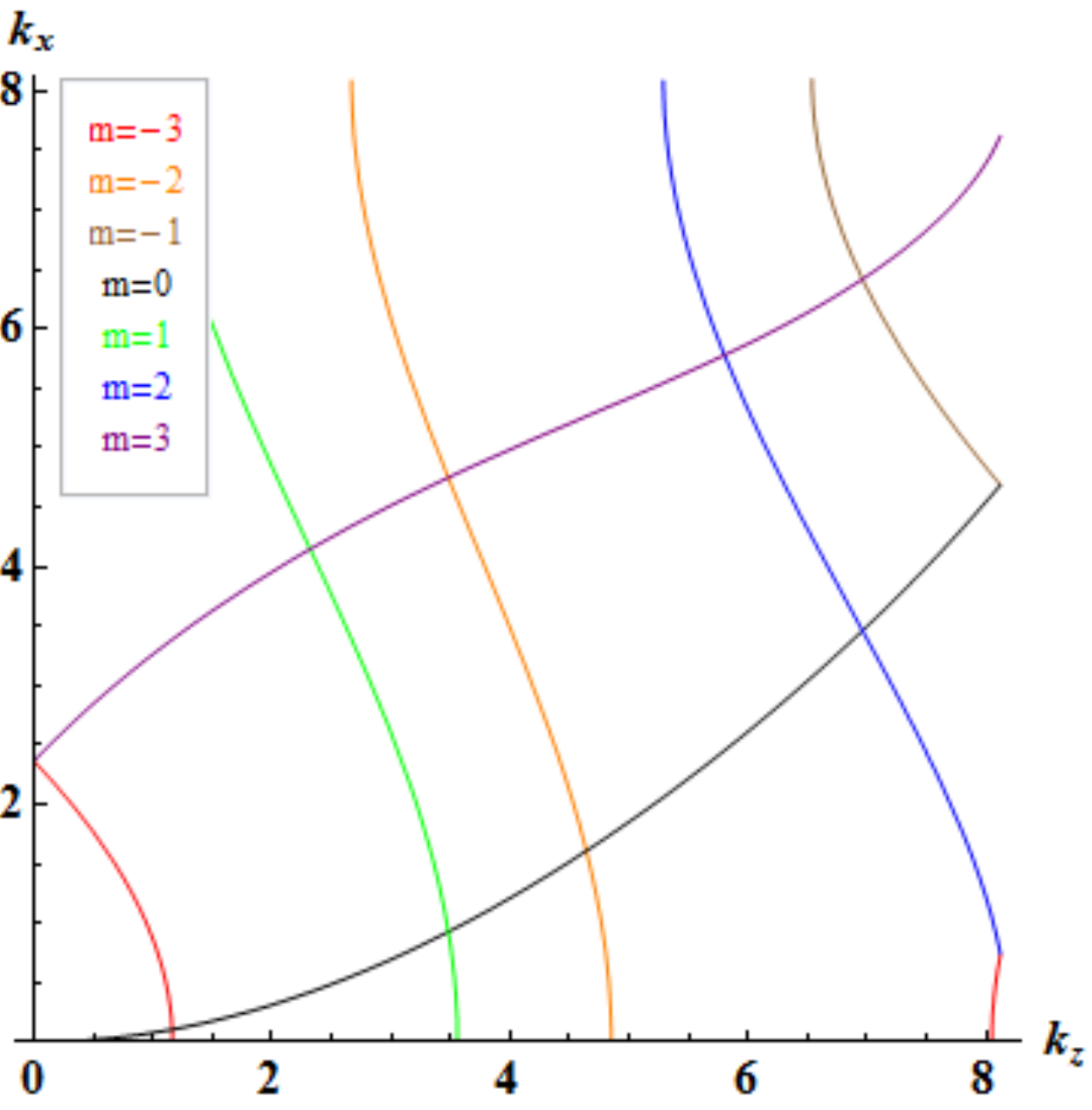}
\par\end{centering}

\caption{\label{fig:Resonance curves}Locations in \textit{k}-space of resonances
between electromagnetic modes and beam modes, $m_{z}=\left[-3,\,3\right]$
for $\frac{\Delta t}{\Delta z}=0.7$ and $\beta=0$. Intersecting
resonance curves occur at different frequencies and, therefore, do
not interact.}
\end{figure}
\clearpage{}
\begin{figure}
\centering{}\includegraphics{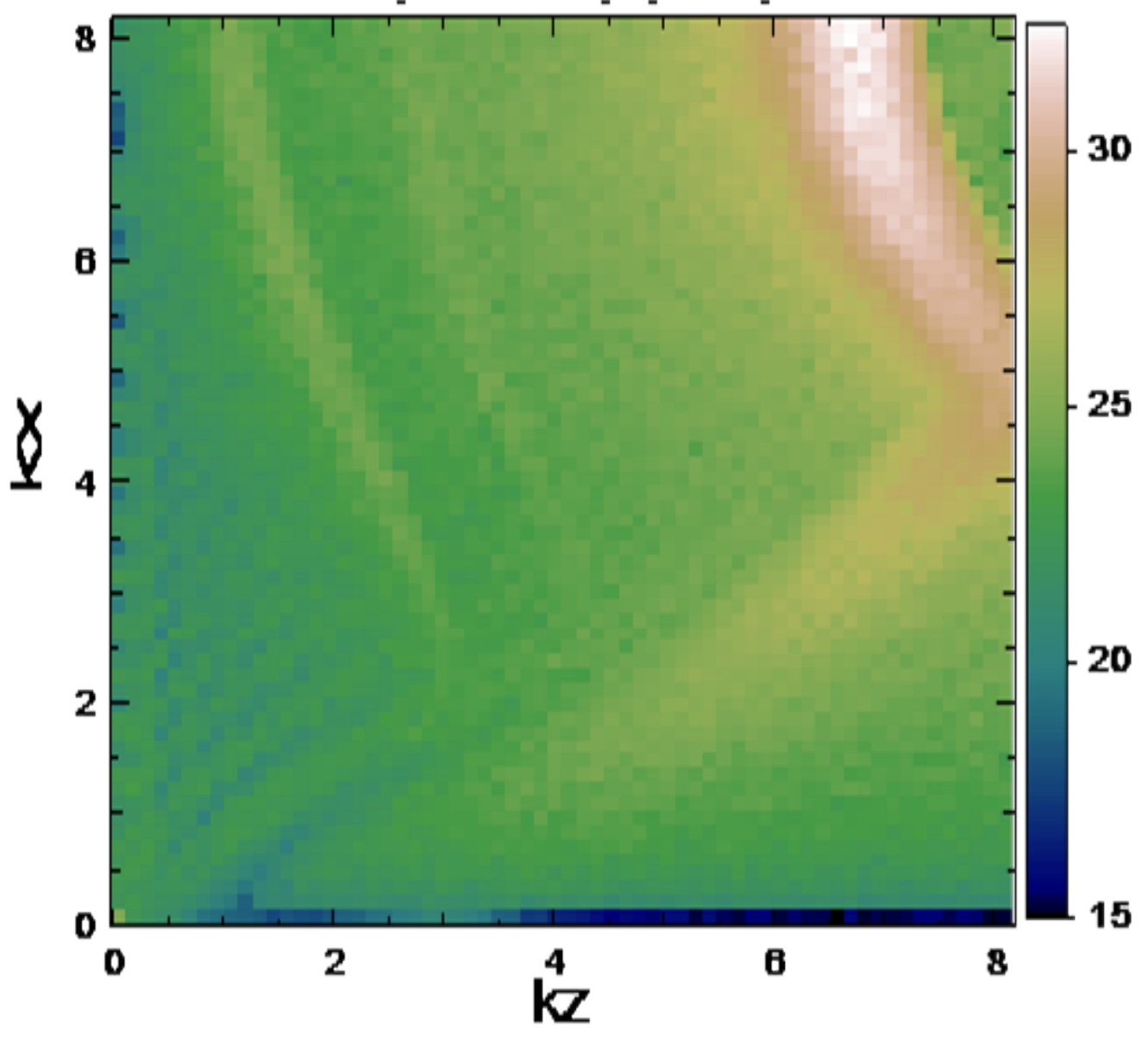}\caption{\label{fig:Ez sample}Fourier-transformed $E_{z}$ (log scale) at
$t=16$ from a WARP simulation with Galerkin field interpolation,
$\frac{\Delta t}{\Delta z}=0.7$, and $\beta=0$.}
\end{figure}
\clearpage{}
\begin{figure}
\begin{centering}
\includegraphics[scale=0.7]{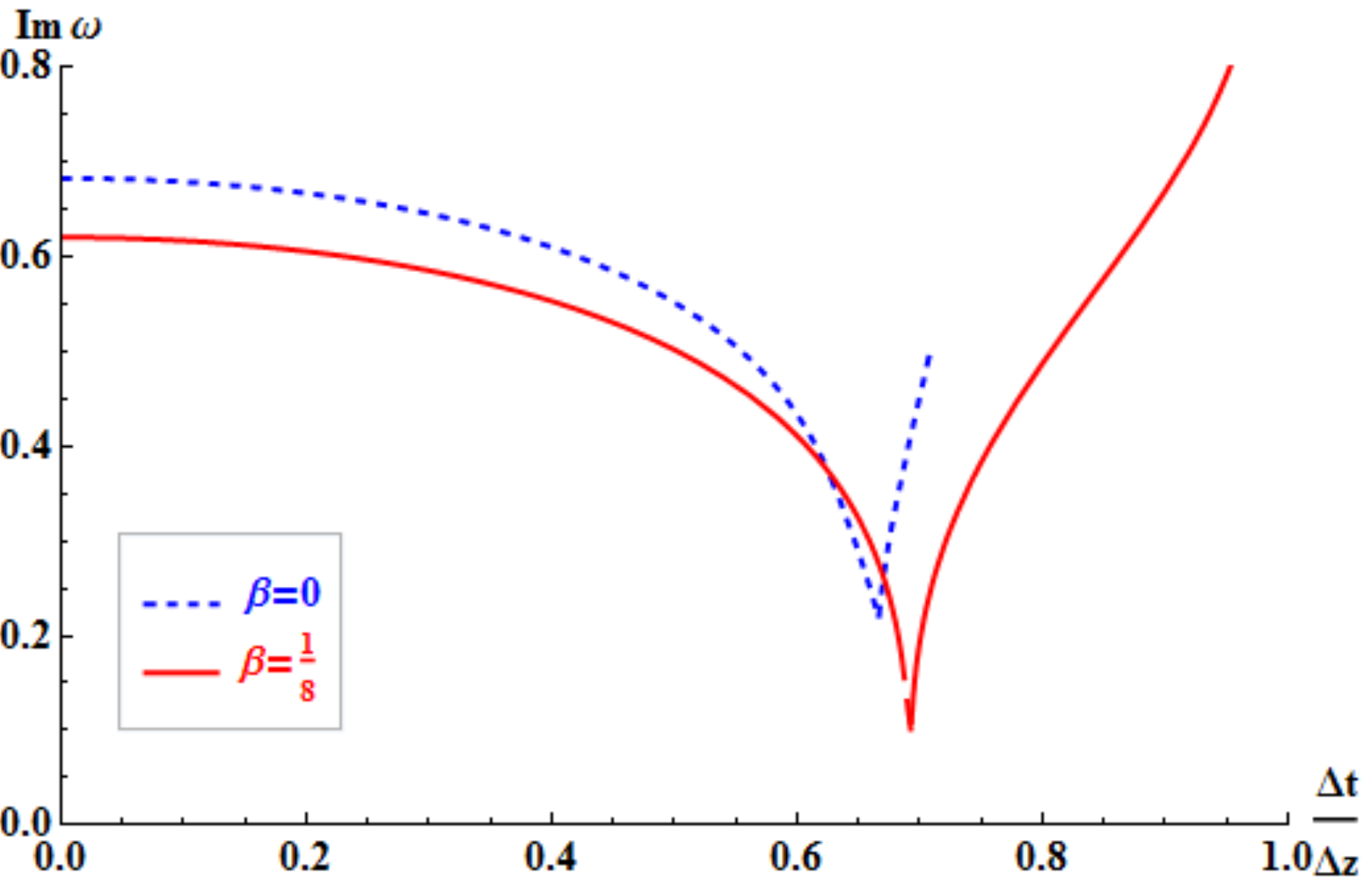}
\par\end{centering}

\caption{\label{fig:Galerkin approx growth}Approximate peak growth rate vs
$\Delta t/\Delta z$ for Galerkin field interpolation with $\beta=0,\nicefrac{1}{8}$. }
\end{figure}
\clearpage{}
\begin{figure}
\begin{centering}
\includegraphics[scale=0.7]{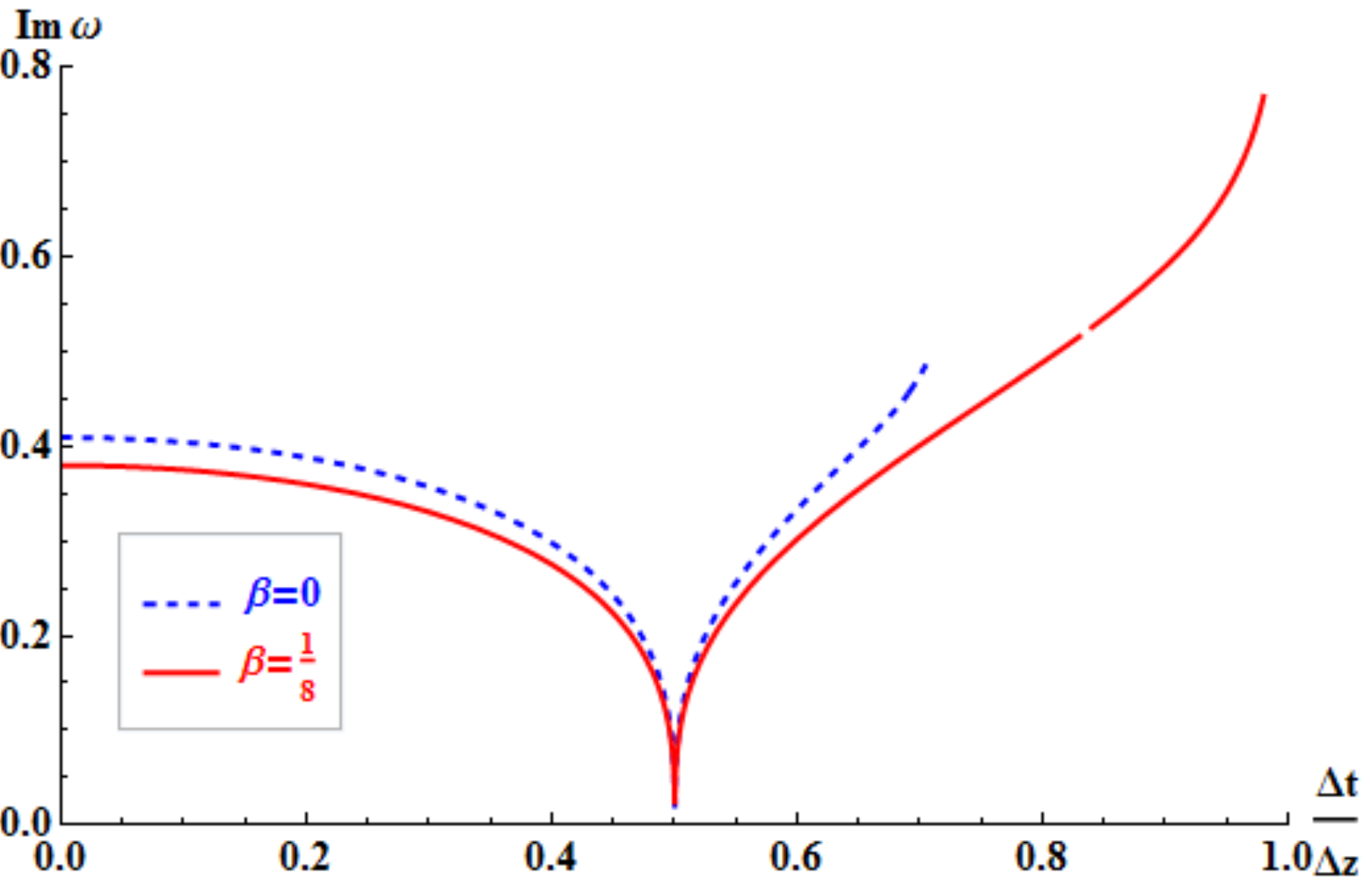}
\par\end{centering}

\caption{\label{fig:Uniform approx growth}Approximate peak growth rate vs
$\Delta t/\Delta z$ for uniform field interpolation with $\beta=0,\nicefrac{1}{8}$.
The growth rate vanishes at $\Delta t=\Delta z/2$ for all values
of $\beta$.}
\end{figure}
\clearpage{}
\begin{figure}
\centering{}\includegraphics[scale=0.8]{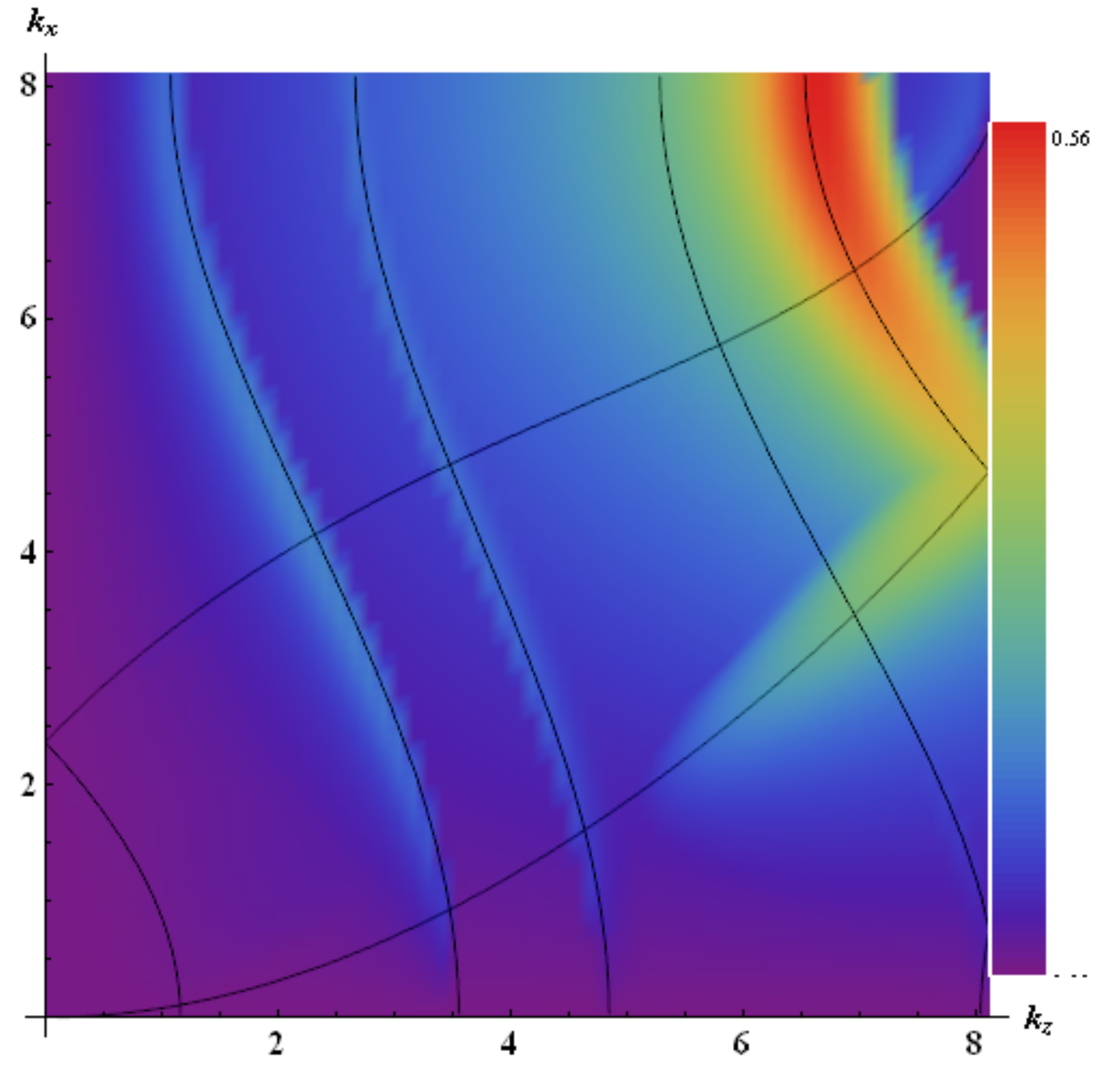}\caption{\label{fig:Numerical growth sample}Instability growth rates calculated
from the numerical dispersion relation for Galerkin field interpolation,
$\frac{\Delta t}{\Delta z}=0.7$, and $\beta=0$. Fig. \ref{fig:Ez sample}
shows corresponding WARP results. Resonance curves are as in Fig.
\ref{fig:Resonance curves}.}
\end{figure}
\clearpage{}
\begin{figure}
\begin{centering}
\includegraphics[scale=0.8]{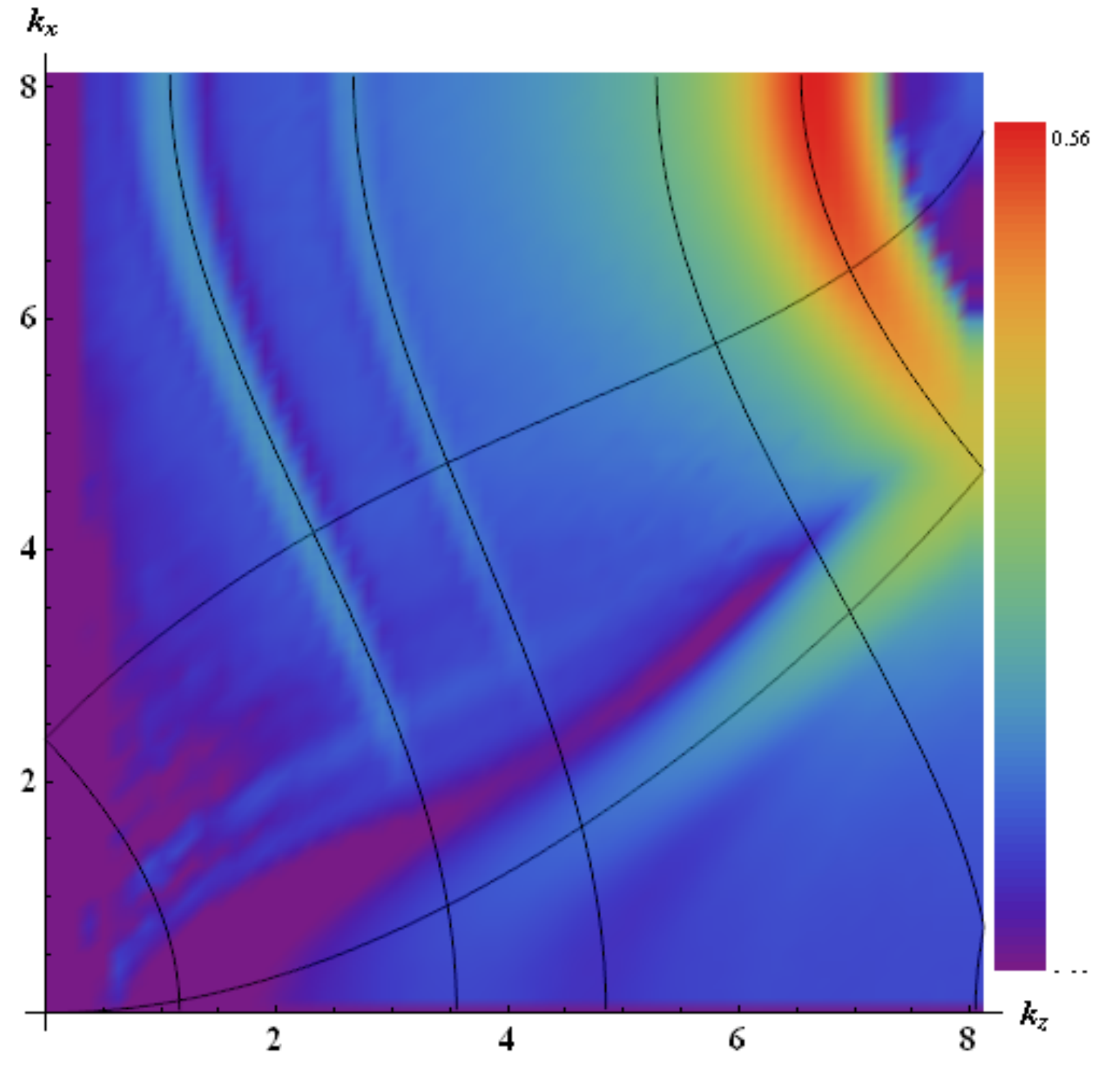}
\par\end{centering}

\caption{\label{fig:Growth rate sample}Instability growth rates for Galerkin
field interpolation, $\frac{\Delta t}{\Delta z}=0.7$, and $\beta=0$,
computed from WARP simulations characterized by Fig. \ref{fig:Ez sample} }
\end{figure}
\clearpage{}
\begin{figure}
\centering{}\includegraphics{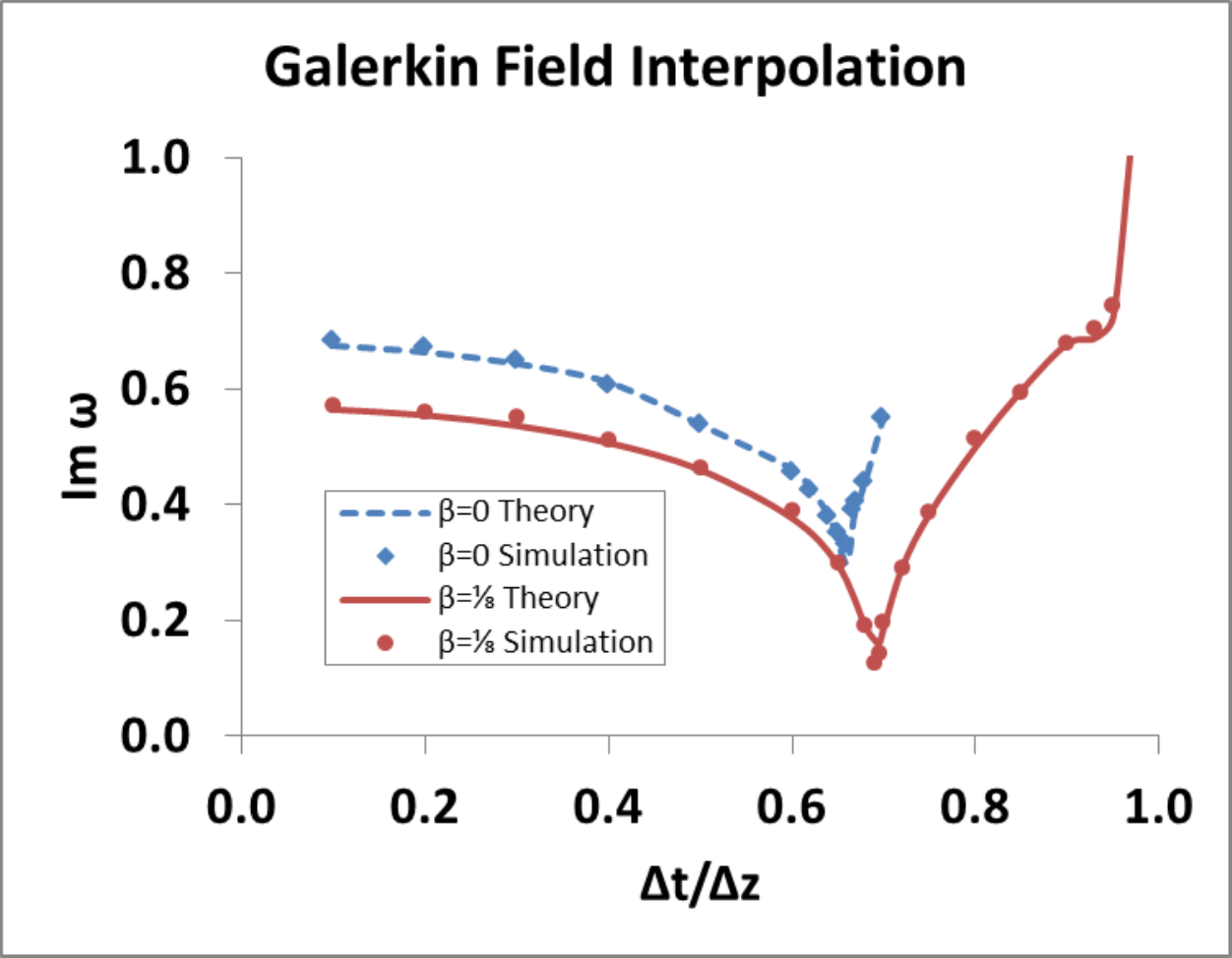}\caption{\label{fig:Galerkin growth scan}Maximum numerical instability growth
rates observed in WARP and calculated from the numerical dispersion
relation for Galerkin field interpolation with $\beta=0,\nicefrac{1}{8}$.}
\end{figure}
\clearpage{}
\begin{figure}
\begin{centering}
\includegraphics{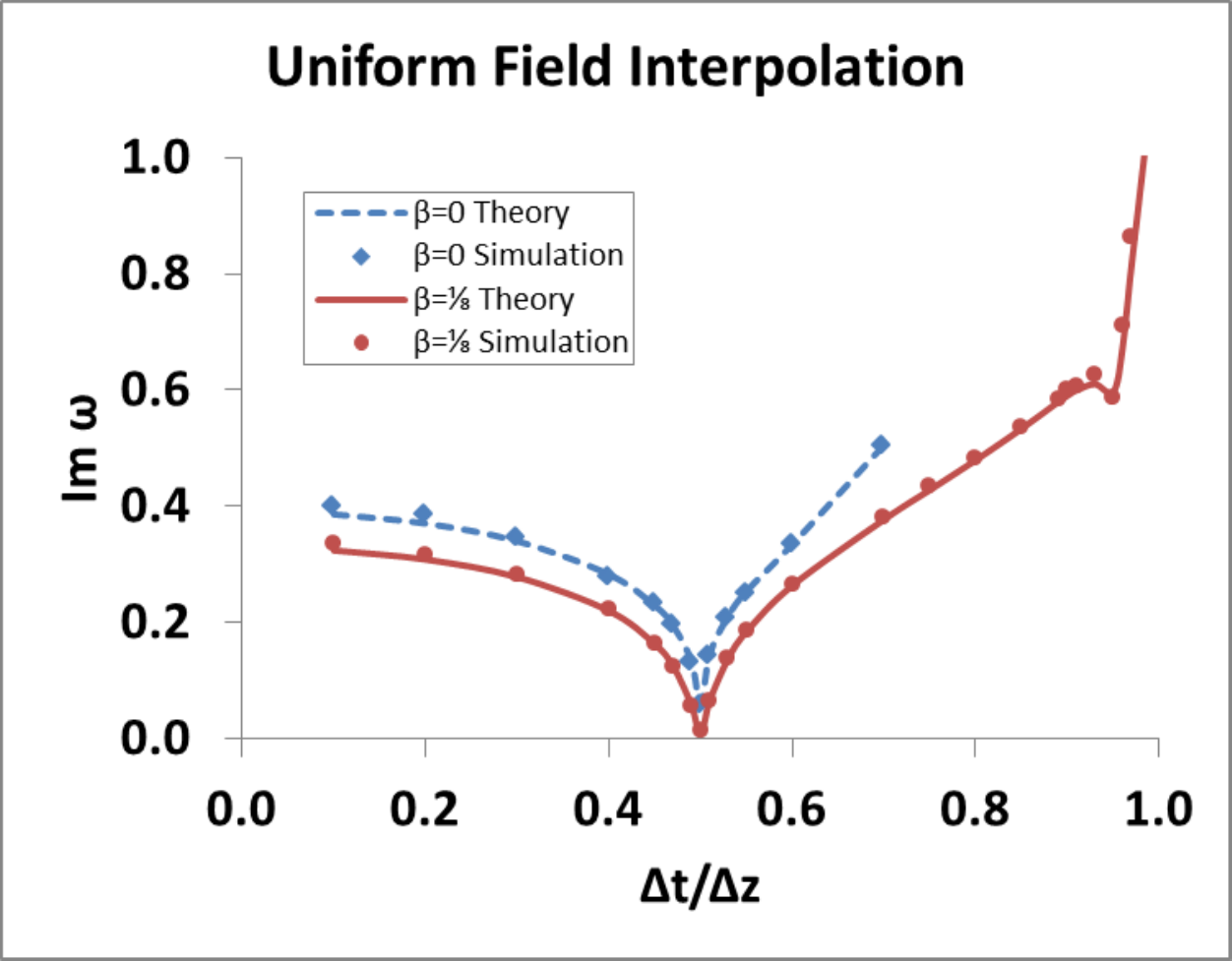}
\par\end{centering}

\caption{\label{fig:Uniform growth scan}Maximum numerical instability growth
rates observed in WARP and calculated from the numerical dispersion
relation for uniform field interpolation with $\beta=0,\nicefrac{1}{8}$.}
\end{figure}
\clearpage{}
\begin{figure}
\begin{centering}
\includegraphics{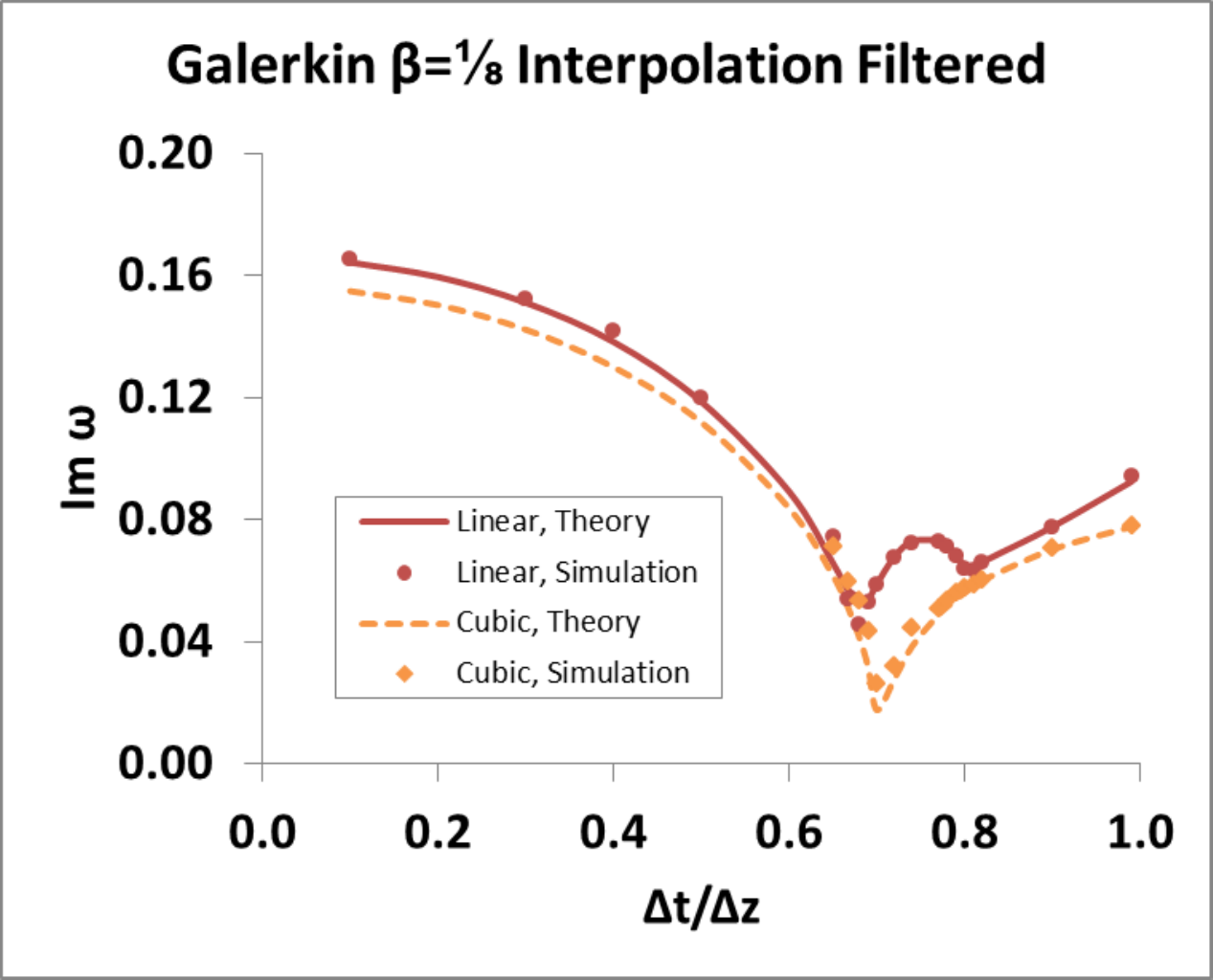}
\par\end{centering}

\caption{\label{fig:Smoothing}Maximum numerical instability growth rates observed
in WARP and calculated from the numerical dispersion relation for
digital filtering as described in Sec. 5, overall linear or cubic
interpolation in \emph{z}, and Galerkin field interpolation with $\beta=\nicefrac{1}{8}$.}
\end{figure}
\clearpage{}
\begin{figure}
\begin{centering}
\includegraphics[scale=0.7]{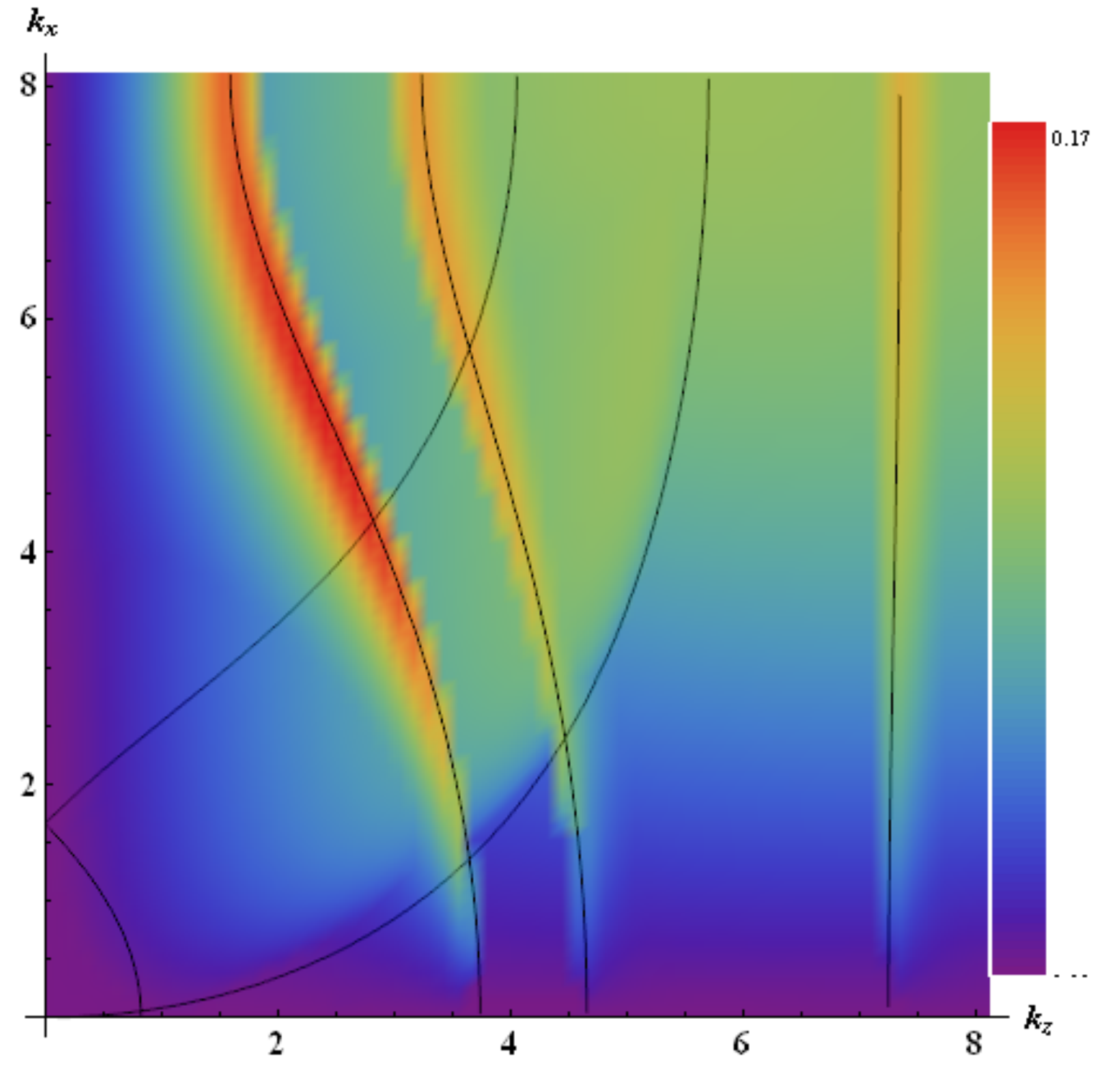}
\par\end{centering}

\caption{\label{fig:DISP multi-alias}Instability growth rates calculated from
the numerical dispersion relation with Galerkin field interpolation,
$\frac{\Delta t}{\Delta z}=0.69$, and $\beta=\nicefrac{1}{8}$.}
\end{figure}
\pagebreak{}
\begin{figure}
\begin{centering}
\includegraphics[scale=0.8]{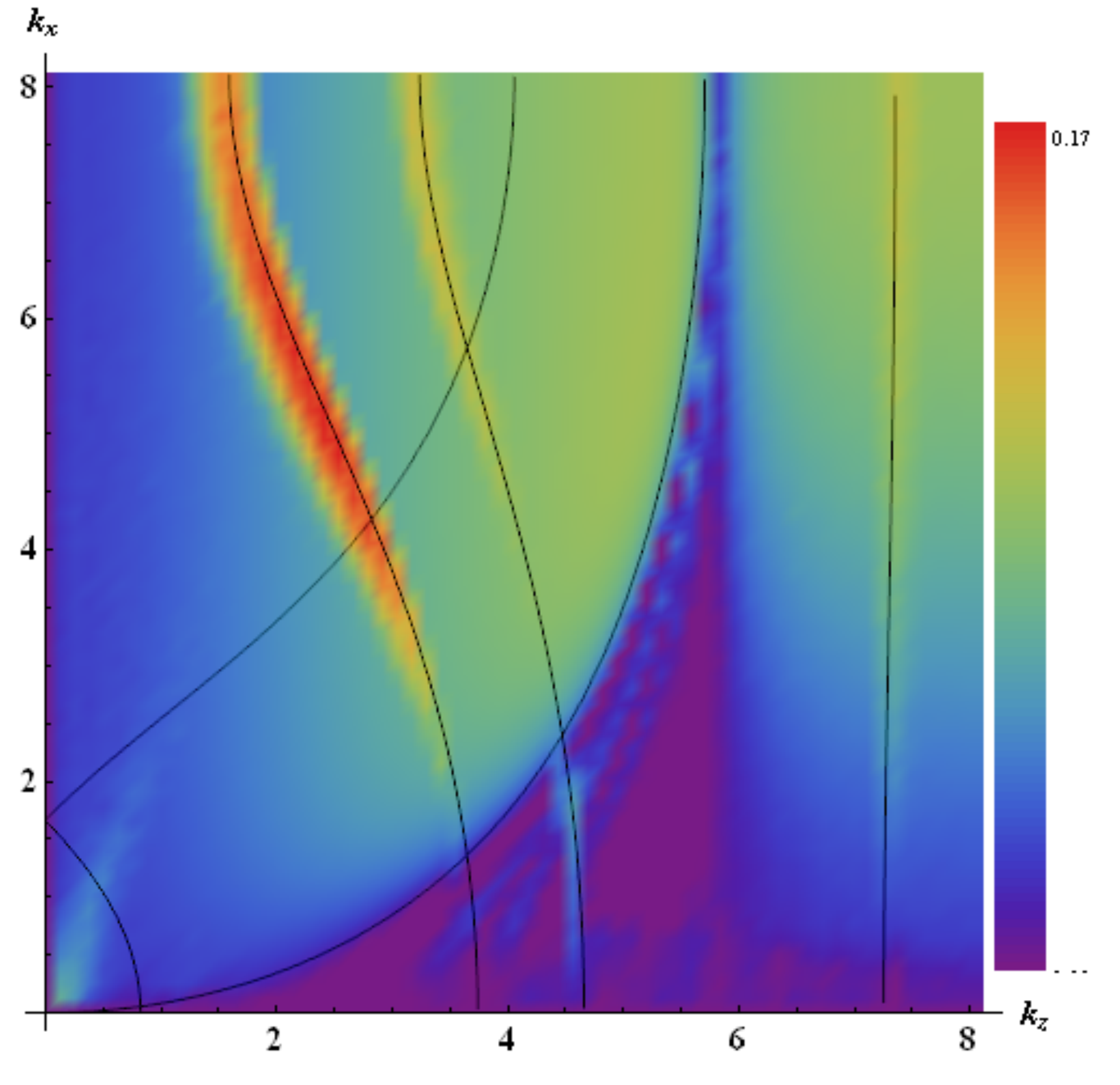}
\par\end{centering}

\caption{\label{fig:WARP multi-alias}Instability growth rates observed in
WARP simulations with Galerkin field interpolation, $\frac{\Delta t}{\Delta z}=0.69$,
and $\beta=\nicefrac{1}{8}$.}
\end{figure}
\clearpage{}
\begin{figure}
\centering{}\includegraphics[scale=0.5]{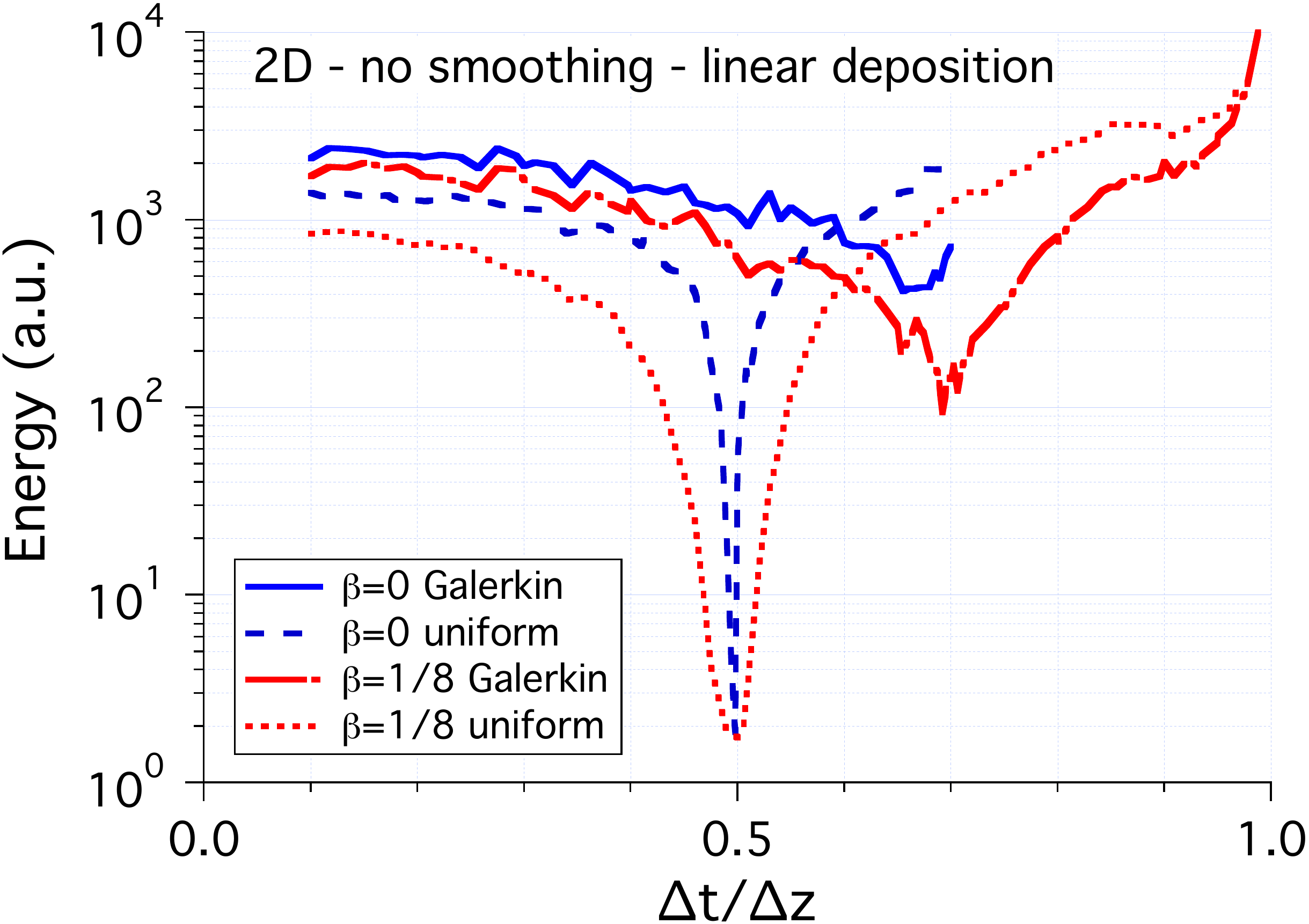}\caption{\label{fig:WARP saturation}Field energy relative to stable reference
level vs $\Delta t/\Delta z$ from two dimensional WARP LPA simulations
at $\gamma$ = 13, using Galerkin and uniform field interpolation
with $\beta=0,\nicefrac{1}{8}$, no filtering, and linear interpolation.}
\end{figure}
\clearpage{}
\begin{figure}
\centering{}\includegraphics[scale=0.5]{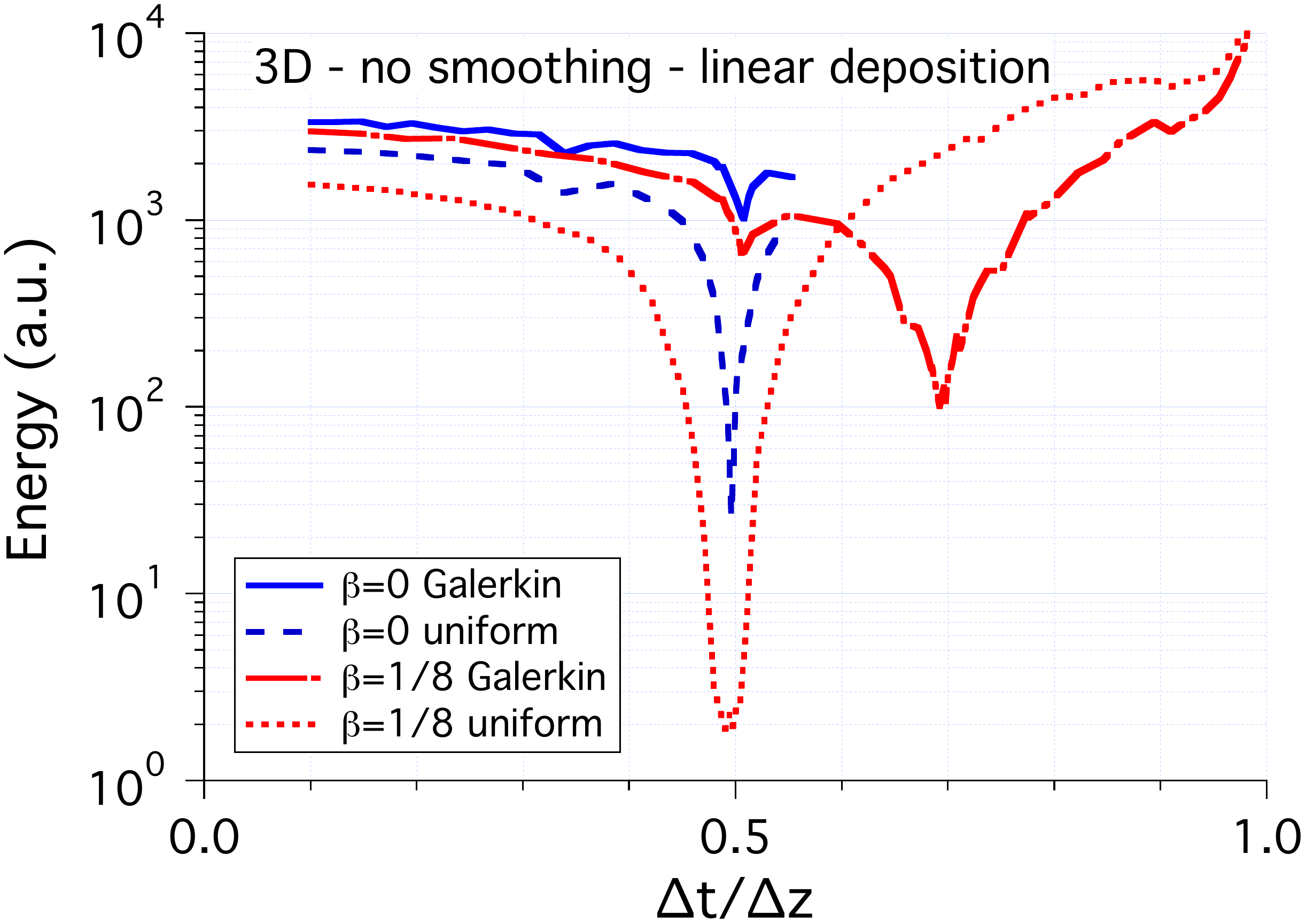}\caption{\label{fig:WARP saturation-3D}Field energy relative to stable reference
level vs $\Delta t/\Delta z$ from three dimensional WARP LPA simulations
at $\gamma$ = 13, using Galerkin and uniform field interpolation
with $\beta=0,\nicefrac{1}{8}$.}
\end{figure}
\clearpage{}
\begin{figure}
\centering{}\includegraphics[scale=0.5]{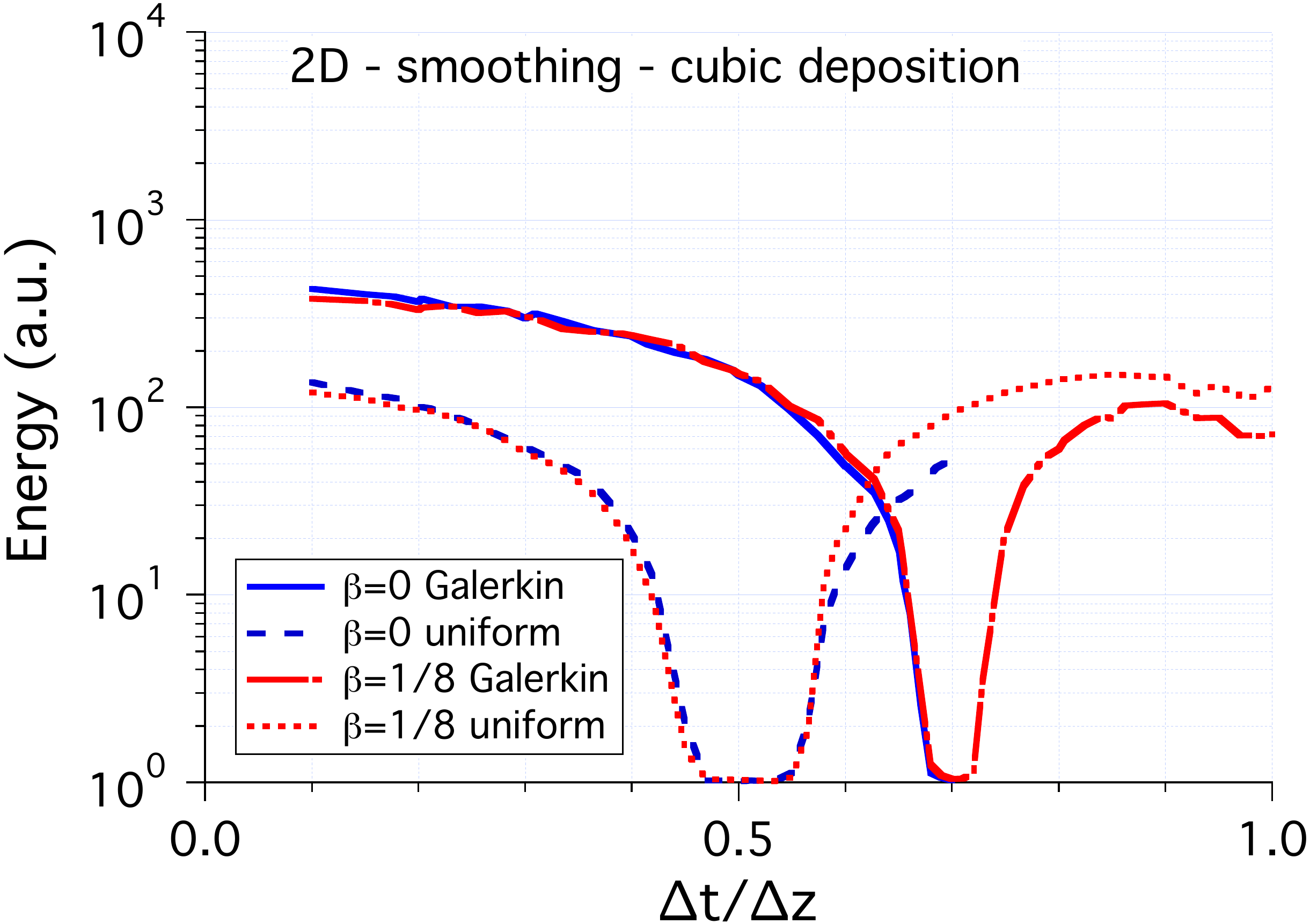}\caption{\label{fig:WARP saturation-2D-smooth}Field energy relative to stable
reference level vs $\Delta t/\Delta z$ from three dimensional WARP
LPA simulations at $\gamma$ = 13, using Galerkin and uniform field
interpolation with $\beta=0,\nicefrac{1}{8}$.}
\end{figure}
\clearpage{}
\begin{figure}
\centering{}\includegraphics[scale=0.5]{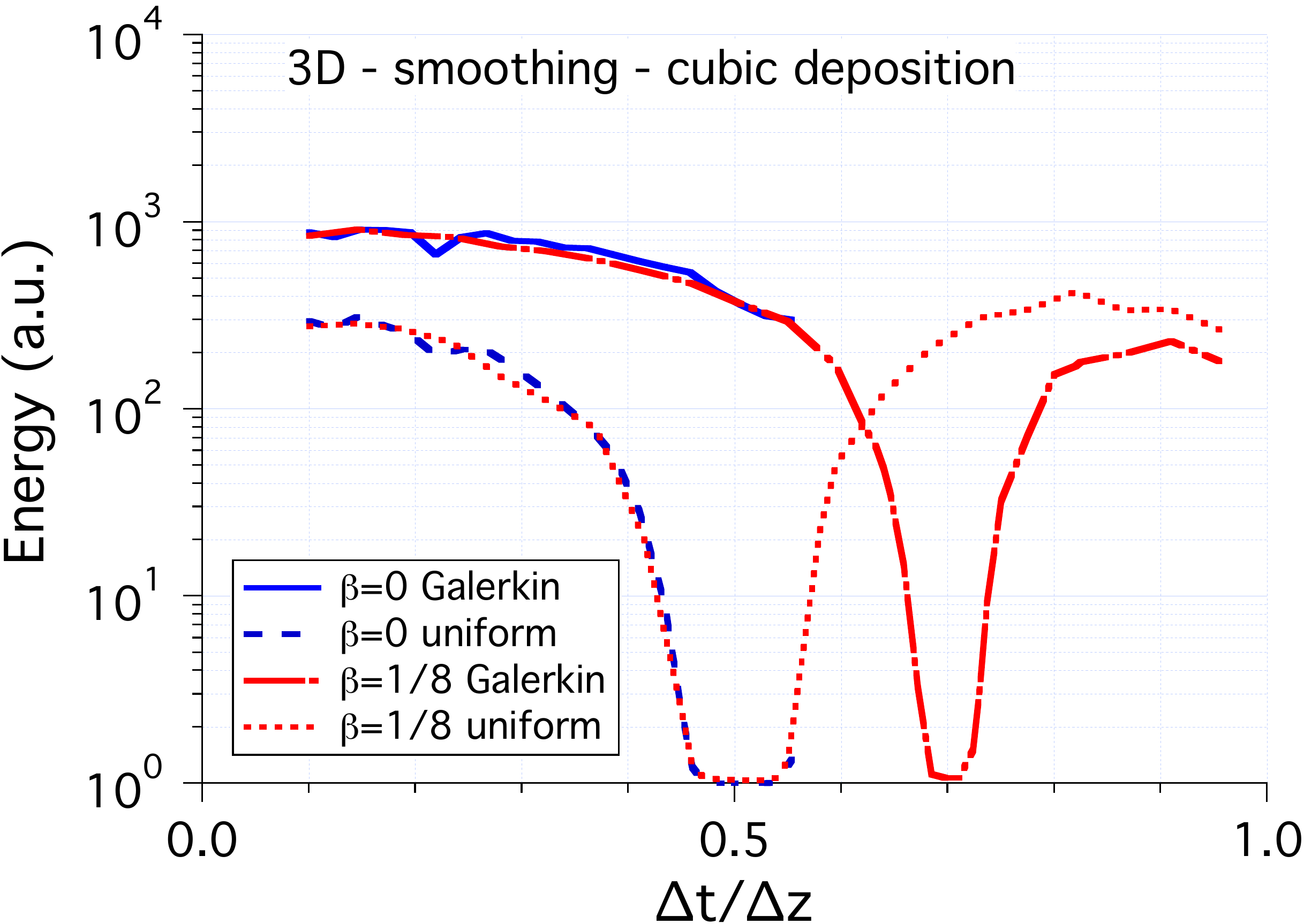}\caption{\label{fig:WARP saturation-3D-smooth}Field energy relative to stable
reference level vs $\Delta t/\Delta z$ from three dimensional WARP
LPA simulations at $\gamma$ = 13, using Galerkin and uniform field
interpolation with $\beta=0,\nicefrac{1}{8}$.}
\end{figure}
\clearpage{}This document was prepared as an account of work sponsored
in part by the United States Government. While this document is believed
to contain correct information, neither the United States Government
nor any agency thereof, nor The Regents of the University of California,
nor any of their employees, nor the authors makes any warranty, express
or implied, or assumes any legal responsibility for the accuracy,
completeness, or usefulness of any information, apparatus, product,
or process disclosed, or represents that its use would not infringe
privately owned rights. Reference herein to any specific commercial
product, process, or service by its trade name, trademark, manufacturer,
or otherwise, does not necessarily constitute or imply its endorsement,
recommendation, or favoring by the United States Government or any
agency thereof, or The Regents of the University of California. The
views and opinions of authors expressed herein do not necessarily
state or reflect those of the United States Government or any agency
thereof or The Regents of the University of California. 
\par\end{center}
\end{document}